\documentclass[default,iicol,10pt]{sn-jnl}

\jyear{2021}%

\theoremstyle{thmstyleone}%
\theoremstyle{thmstyletwo}%

\theoremstyle{thmstylethree}%

\begin{document}

\title[Article Title]{A framework for crossover of scaling law as a self-similar solution : dynamical impact of viscoelastic board}


\author*[1,2]{\fnm{Hirokazu} \sur{Maruoka}}\email{hmaruoka1987@gmail.com}\email{hirokazu.maruoka@yukawa.kyoto-u.ac.jp}

\affil*[1]{\orgdiv{Deep-Sea Nanoscience Research Group, Research Center for Bioscience and Nanoscience, Research Institute for Marine Resouce Utilization}, \orgname{Japan Agency for Marine-Earth Science and Technology (JAMSTEC)}, \orgaddress{\street{2-15 Natsushima-cho}, \city{Yokosuka}, \postcode{237-0061}, \state{Kanagawa}, \country{Japan}}}

\affil*[2]{\orgdiv{Advanced Statistical Dynamics}, \orgname{Yukawa Institute for Theoretical Physics, Kyoto University)}, \orgaddress{\street{Kitashirakawa Oiwakecho}, \city{Sakyo-ku}, \postcode{606-8502}, \state{Kyoto}, \country{Japan}}}


\abstract{In this paper, a new framework for crossover of scaling law is proposed: a crossover of scaling law can be described by a self-similar solution. A crossover emerges as a result of the interference from similarity parameters of the higher class of the self-similarity. This framework was verified for the dynamical impact of a solid sphere onto a viscoelastic board. All the physical factors including the size of spheres and the impact of velocity are successfully summarized using primal dimensionless numbers which construct a self-similar solution of the second kind, which represents the balance between dynamical elements involved in the problem. The self-similar solution gives two different scaling laws by the perturbation method describing the crossover. These theoretical predictions are compared with experimental results to show good agreement. It was suggested that a hierarchical structure of similarity plays a fundamental role in crossover, which offers a fundamental insight into self-similarity in general.}

\maketitle 

\section{Introduction}\label{sec1}

\textquotedblleft Scaling never appears by accident\textquotedblright \cite{Barenblatt2003}. Scaling law is the representation of physical law, which is expressed by a power-functional relation between physical parameters (e.g., Boyle's law is the inverse-proportional relation between pressure and the volume $P \sim V^{-1}$)
\begin{equation}
y  =  At^{\alpha}
\label{eq:E1a}
\end{equation} 
in which $y$, $t$ are physical parameters, $A$ is a prefactor and $\alpha$ is a power exponent. It is a quite general and basic concept in physics. It enables us to connect theory with experiment as theoretical verification is generally performed through the reference of a scaling relation obtained by experimental observations\cite{deGennes1979}. On the other hand, one observes the case in which a scaling law transforms to another scaling law in different scale of physical variables, $y=At^{\alpha} \to Bt^{\gamma}$, which we call a {\it crossover of scaling law}, in a wide variety of fields: the mechanics of continua\cite{Okumura2011,Murano2020}, soft matter\cite{Berry}, quantum physics\cite{Vasseur2013}, critical phenomena\cite{Lujiten1997,Lubeck2003} and so on. Understanding such phenomena are useful for application and biology as it is expected that they can be associated with the invention of functional materials\cite{Parnell2019,Raphael2021}, and may play an important role in biological functions\cite{Weizsacker1950, Bhushan, Krohs}. A crossover of scaling law can be formalized as {\it the process of transition of scaling law by the continuous change of a scale parameter}. However, the studies of crossovers generally focus on the extreme limit of each scaling law independently. As a result, they failed to formalize it as a continuous process and understand the mechanism behind the crossover.

The appearance of a stable scaling law can be understood as an {\it intermediate asymptotic}\cite{InterAsym, BarenInter,Barenblatt1996,Barenblatt2014,Barenblatt1972,Goldenfeld,Goldenfeld1992,Oono2013,Benzaquen2013,Baumchen2013,Maruoka2019}, which is defined as an asymptotic representation of a function valid in a certain scale range. Barenblatt has formalized the idealization of physical theory based on dimensional analysis. Dimensional analysis gives a self-similar solution of which variables are dimensionless numbers consisting of the physical quantities involved in the phenomena. Considering the dimensions of parameters, the scaling law of Eq.~(\ref{eq:E1a}) can be transformed to $\Pi = y/At^{\alpha}$. Later the dependence of $\Pi$ on other dimensionless numbers, say $\theta = t^{\beta}$/x, is investigated to obtain $\Pi = \Phi (\theta)$. If the dimensionless function $\Phi$ converges to a finite limit, $\Phi \rightarrow {\rm const}$ as $\theta \rightarrow 0$, which corresponds to {\it complete similarity}\cite{CompSim}, and if a single dimensionless number remains, an intermediate asymptotic is obtained, which results in the scaling law corresponding to Eq.~(\ref{eq:E1a}). Note that it is {\it locally} valid in the range in which its asymptotic is maintained in this case; it is $y = At^{\alpha}~( \theta \ll 1,  0 < t \ll x^{1/\beta})$\cite{InterAsym}. This formalization can facilitate our understanding of the idealization in physics through dimensionless numbers. However, his theory is limited to the case of a single scaling law and has not been extended to a crossover of scaling law. 

In this paper, I develop Barenblatt's idea for a crossover of scaling law. In terms of the concept of the intermediate asymptotic, it is expected that a crossover must correspond to a breakdown of its idealization. As previously mentioned, a single scaling law is obtained when its dimensionless function converges to a finite limit, $\Phi = {\rm const}$. Conversely speaking, the incomplete convergence of a dimensionless function, $\Phi \neq {\rm const}$, namely the interference of another dimensionless number may generate a crossover of scaling law. Therefore, $\Phi$ is a mechanism for changing the intermediate asymptotics to another. I will demonstrate that we can understand crossover by such a framework. If we find $\Phi$, we can describe the crossover as a continuous process.

In this study, on the dynamical impact of a solid sphere onto a viscoelastic board, I show how the Maxwell model can become a mechanism to generate a crossover of scaling law, and how it constitutes a self-similar solution in which two dimensionless variables are related, which corresponds to the aforementioned framework. The Maxwell model is frequently applied to the various phenomena in which the behaviors change on the different time-scale, such as earthquake\cite{Suito,Agata}, and fracture\cite{Persson,SakuOku}. In the context of contact mechanics\cite{Hertz,Johnson, Goryacheva}, in which features of contact are drastically changed\cite{Carpick,Kogut} depending on the form of contact, the viscoelastic materials have provided interesting materials\cite{Hunter,Hertzsch,Brilliantov}. It is recognized that viscoelasticity plays an important role in the adhesion of interface and closing or opening crack between surface\cite{Persson}. In this work, I focus on the viscoelastic behavior derived from bulk properties by eliminating the adhesion contribution by dusting the surface. I will show that a new scaling law appears in the viscoelastic regime, and that there exists a self-similar solution which governs this crossover of scaling law, which corresponds to a self-similar solution of the second kind. 

\section{A framework for a crossover of scaling law}\label{sec1}

\begin{figure}[h] 
\begin{center} 
\includegraphics[width=1\columnwidth]{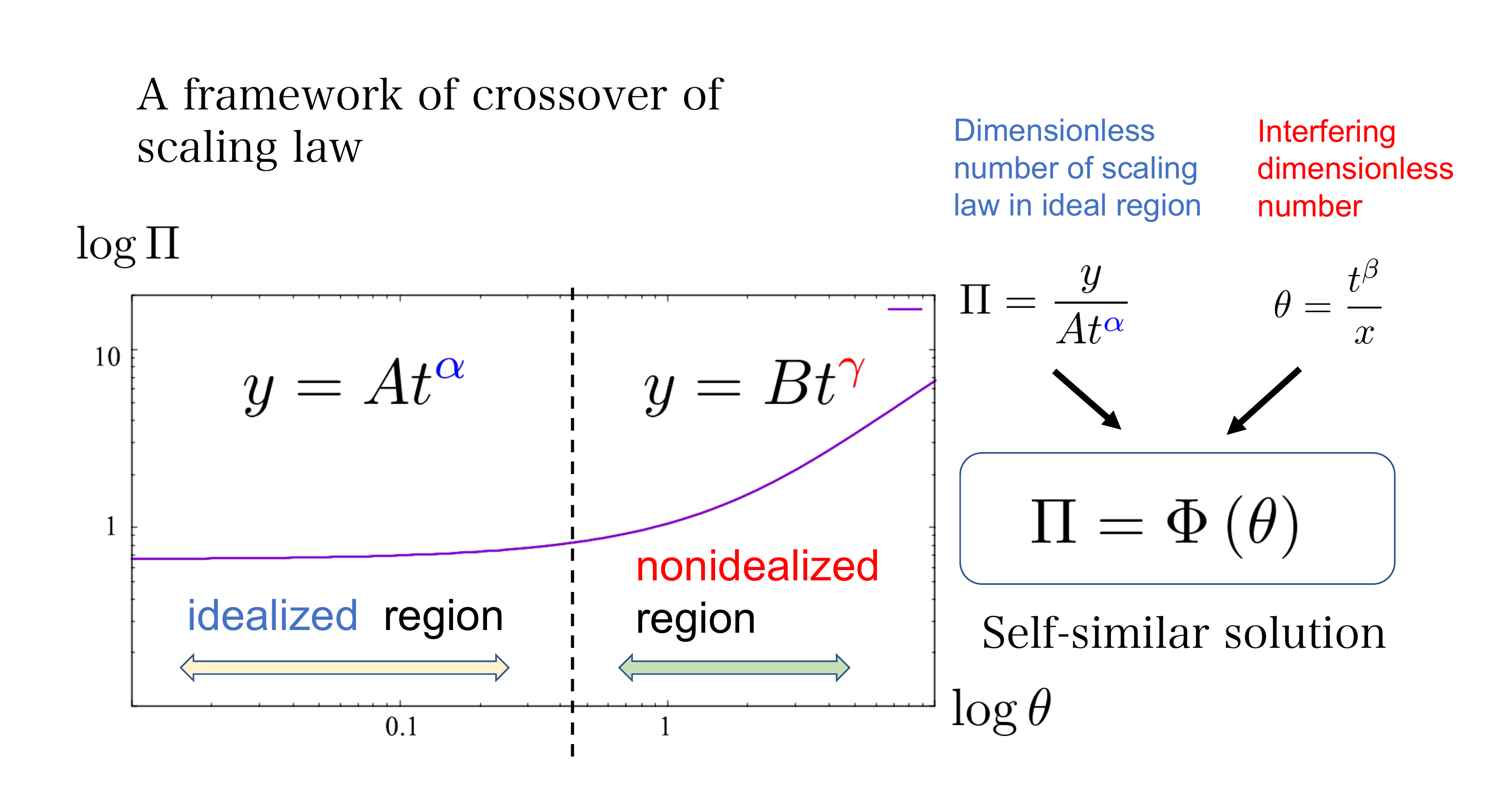}
\caption{(Color online) A framework of crossover of scaling law. By introducing a dimensionless number of a scaling law in an idealized region as $\Pi =\frac{y}{At^{\alpha}}$, one can describe how the scaling law is changed by another dimensionless number $\theta$ as a self-similar solution, $\Pi = \Phi \left( \theta\right)$. $\theta$ is composed of the interfering physical quantity $x$.  }
\label{fig:F1a}
\end{center}
\end{figure} 

A framework to describe a crossover of scaling law is explained as follows (Fig. \ref{fig:F1a}). A crossover of scaling law can be understood as the process in which a certain scaling law $y = A t^{\alpha}$ in an idealized region is changed by the interference of another physical quantity $x$ to obtain another scaling law $y= B t^{\gamma}$ in the nonidealized region. In terms of self-similarity, this process can be described by the variation of a dimensionless number $\Pi$ by another dimensionless number $\theta$ where $\Pi$ is composed of the scaling law as $\Pi = \frac{y}{At^\alpha}$ and $\theta$ is composed of an interfering physical quantity $x$ as $\theta = \frac{t^{\beta}}{x}$. This relation forms a self-similar solution, $\Pi = \Phi \left( \theta \right)$. If $\Phi$ converges to a finite limit as $\theta \to 0$, namely $\Pi =  \frac{y}{At^\alpha}= 1$ as $\theta \ll 1$, then $y = A t^{\alpha}$ appears as an intermediate asymptotic in the range of $\theta \ll 1$. In this region, $x$ does not interfere with the scaling law. However, out of this idealized region, $\theta$ interferes with the scaling law, then $\Pi$ is no more constant, $\Phi \neq {\rm const}$, then the scaling law transforms into another scaling law. Thus, two scaling laws are integrated by the self-similar solution. The self-similar solution describes the scale-dependence of dimensionless numbers that form scaling laws.

Note that $\Phi$ must satisfy the convergence to a finite limit as $\theta \rightarrow 0$ to recover the scaling law in the idealized region. This process can be considered as the extension of the scaling law in an idealized region to another nonidealized region. It can be a natural procedure for the formalization of problems.

Considering the framework of crossover of scaling law, one can construct such dependence by the following procedure:
\begin{itemize}
      \item Step 1: start from the scaling law valid in a certain scale region, e.g. Eq. (\ref{eq:E1a}) $y = A t^{\alpha}$.
      \item Step 2: define the dimensionless number composed of the scaling law, $\Pi = \frac{y}{At^{\alpha}}$.
      \item Step 3: construct a self-similar solution by identifying an interfering dimensionless number, $\Pi = \Phi \left( \theta \right)$.
\end{itemize}

In this paper, I deal with a crossover of scaling law which is caused by the interference of viscosity. The scaling law of the idealized region is the Chastel-Gondret-Mongruel (CGM) solution which is valid for elastic impacts. Here I attempt to construct a self-similar solution to describe the crossover of scaling law based on the CGR solution and Maxwell Viscoelastic Foundation model.

\section{Experiment}

\begin{figure}[h!]
\includegraphics[width=1\columnwidth]{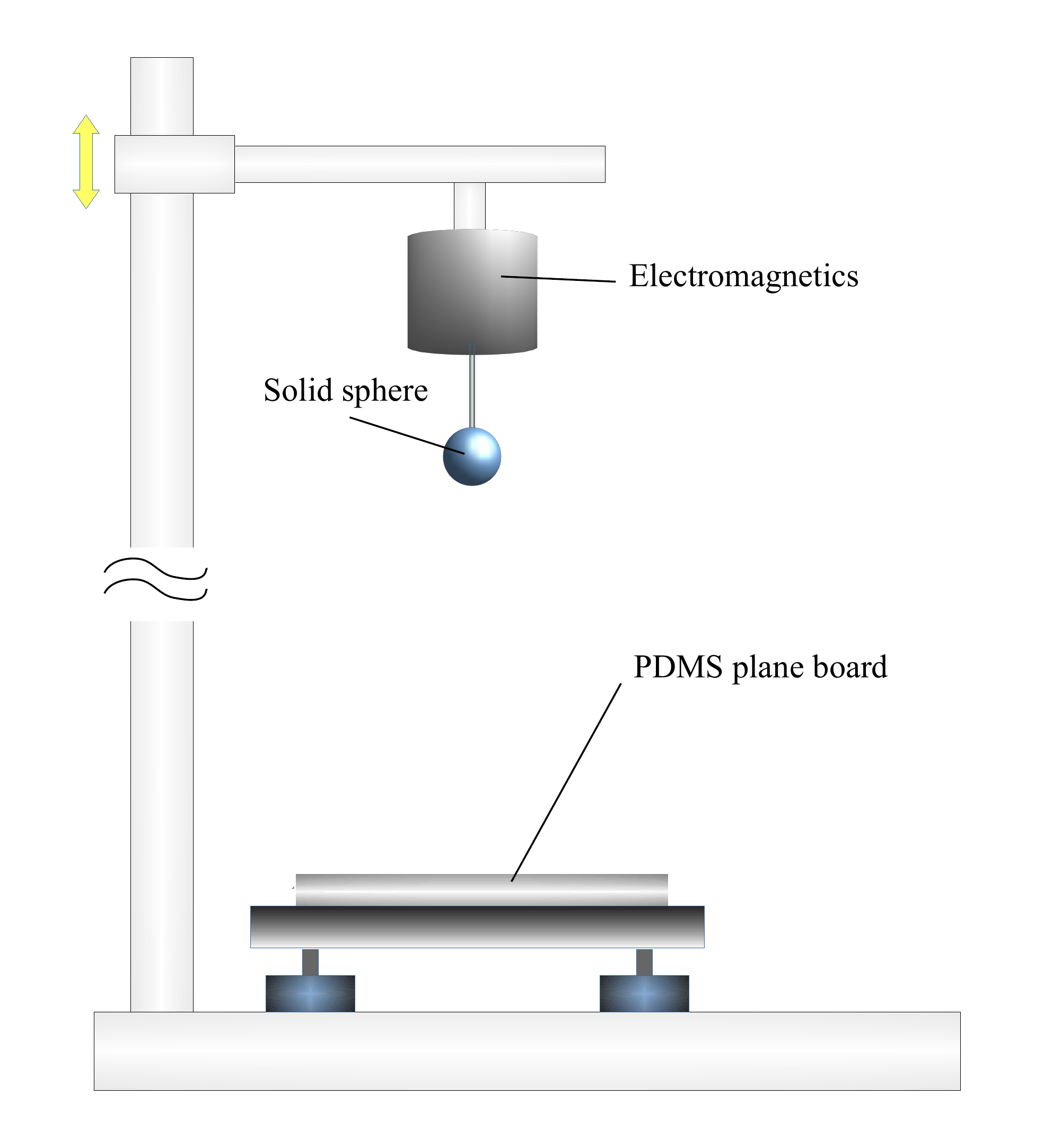}
\caption{(Color online) Sketch of experimental setup. The solid ball is suspended by an electromagnet. The velocity of impact is adjusted by changing the height of the part suspending the sphere. The sphere (R = 3.0, 4.0, 5.0, 6.0, 7.0 and 8.0 mm, $\rho = {\rm 7800~kg \cdot m^{-3}}$) is dropped onto the PDMS plane board ($\phi$ = 1, $h$ = 6.4 mm) by turning off the electromagnet. \label{fig:FS1}}
\end{figure}

\begin{figure}[h!]
  \includegraphics[width=1\columnwidth]{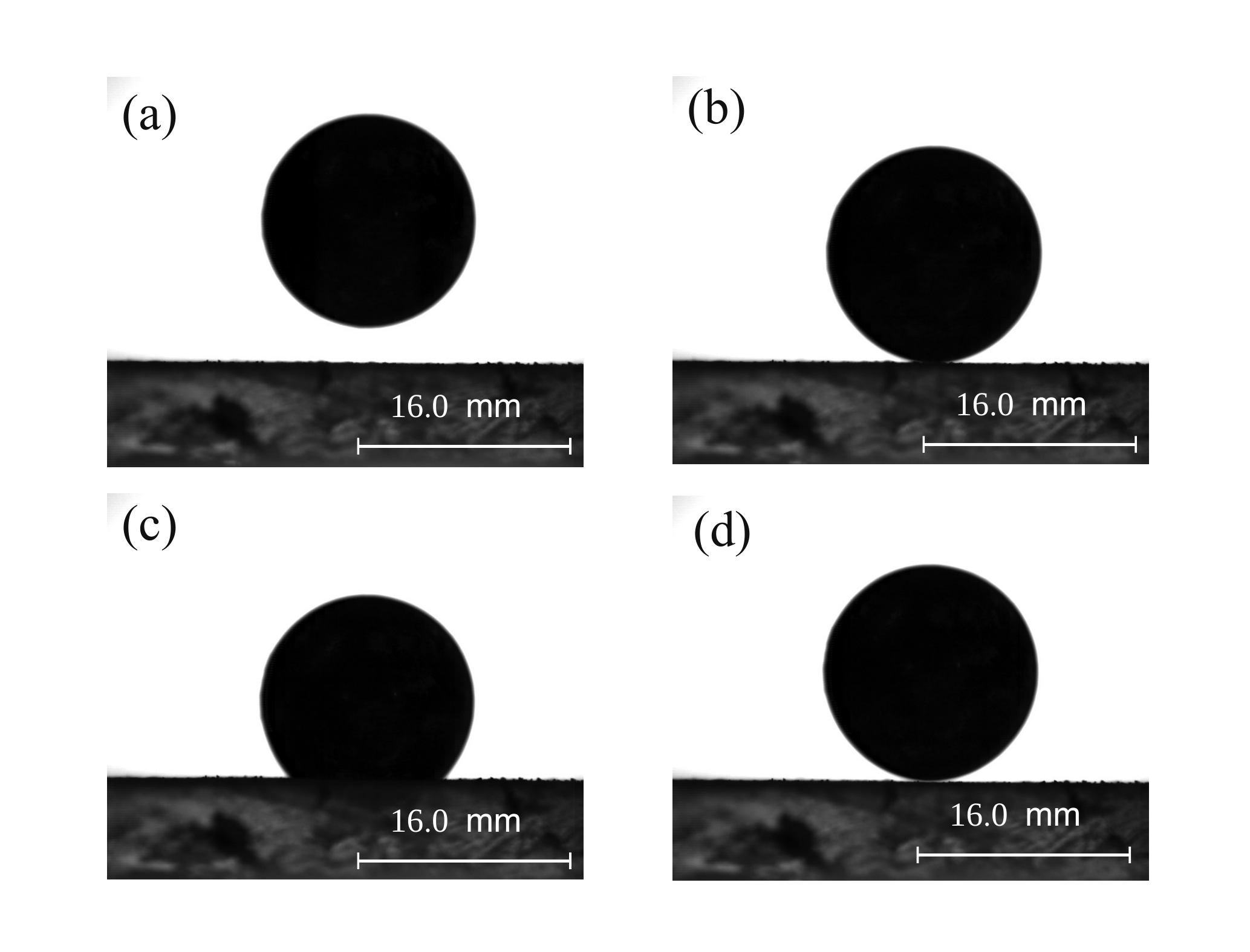}
  \caption{(Color online) Example images for the dynamical impact of a sphere ($R={\rm 8.0~mm}$) onto the viscoelastic board at $v_i={\rm 372~mm/s}$ with the frame rate for 10000 images per second and the resolution of 768$\times$768 pixels. (a) The image before impact. (b) The moment of contact. (c) The moment of maximum deformation at $t={\rm 9~ms}$ after contact. (d) The release of the sphere from the board after contact. The movie is uploaded as Supplemental Material [].  \label{fig:FS3}}
\end{figure}

The experiments have been performed using a viscoelastic board made of polydimethylsiloxane (PDMS)  (Fig. \ref{fig:FS1}). The PDMS (${\rm SILPOT}^{{\rm TM}}$ 184 W/C, DOW) board was prepared by mixing a curing agent and base by a proportion of 1 : 50 and then pour into the mold. After leaving for 3 hr 30 min at 60 ${\rm {}^{\circ}C}$, the board was solidified with a thickness of $h = {\rm 6.4~mm}$, a fraction of contact $\phi=1$, an elastic modulus $E \simeq 0.77~{\rm MPa}$ and a viscous coefficient $\mu = 127 ~ {\rm Pa\cdot s}$. The elastic modulus and viscous coefficient were estimated by fitting the experimental data points, which are used in Figure 6. The prepared PDMS board had a viscosity and smoothness that meant the ball did not rebound by simply dropping the ball due to the effect of adhesion to the surface. To eliminate such an adhesion effect, 
the surface of PDMS board was dusted with chalk powder. The metallic ball (Tsubaki Nakashima co., ltd.,  SUJ2) was suspended by an electromagnet (ESCO Co.,Ltd., EA984CM-1) of which magnet force is controlled. Once the ball is released from the electromagnet, it starts to free fall and collides with the PDMS board (Fig.~\ref{fig:FS3})\cite{SMExp}. After the ball contacted the board, the ball reaches a maximum deformation $\delta_m$ and then rebounds to take off from the board. 

The processes are observed using a high-speed camera (FASTCAM SA1.1, $768 \times 768~{\rm pixel}$, $10000~{\rm fps}$). The size of sphere $R$, varies with a value of 3.0, 4.0, 5.0, 6.0, 7.0 and 8.0 mm, of which the density is taken as $\rho = {\rm 7800~kg \cdot m^{-3}}$. The collision experiments were performed 6 times for each set of dropping. The information of velocity, maximum deformation, contact time, etc. were extracted from the movies by image analysis programmed with Python using Open CV. These numerical estimations were used to calculate the relevant dimensionless numbers.

\section{Maxwell viscoelastic foundation model}

\begin{figure}[h] 
\begin{center} 
\includegraphics[width=6.6cm]{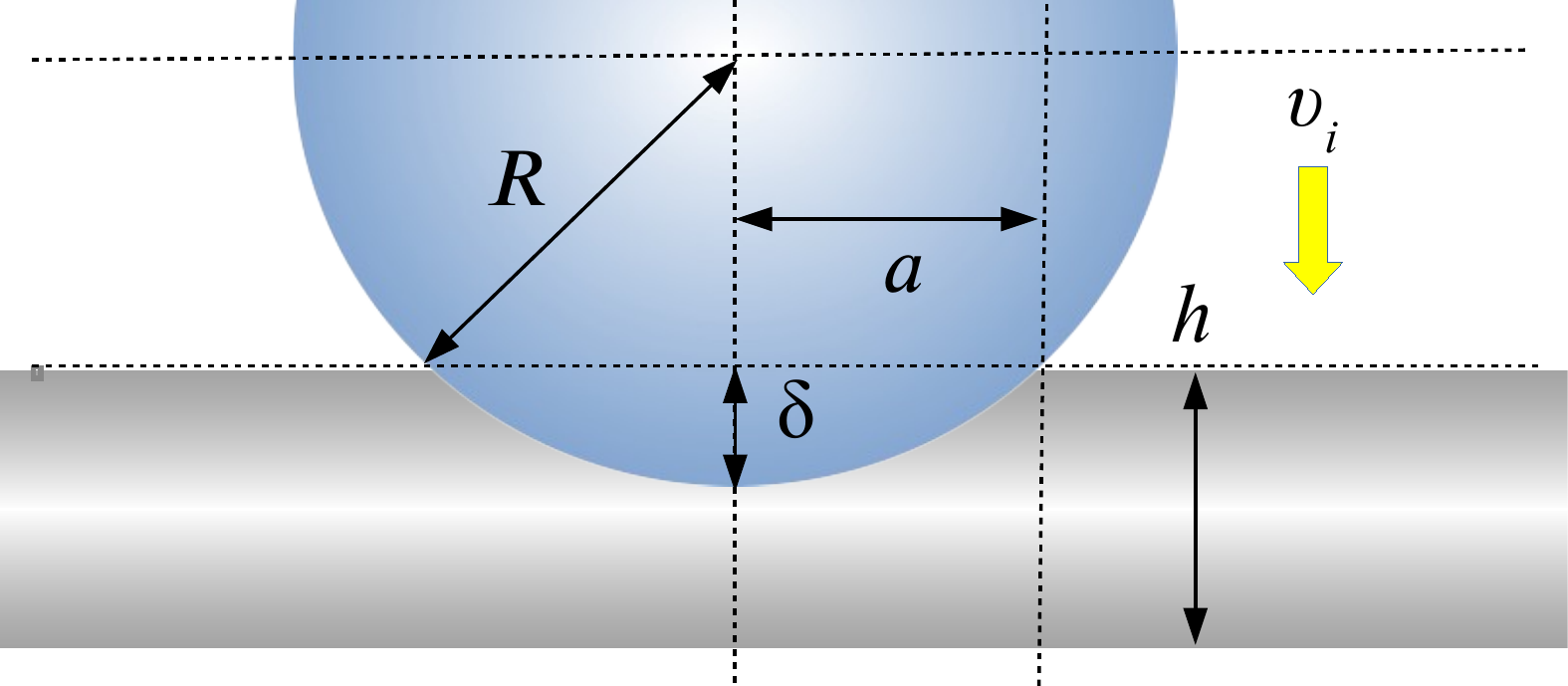}
\caption{(Color online) The geometrical parameters involved in the collision between a viscoelastic board, with a thickness of $h$, and a solid sphere, with its radius $R$ in the impact-velocity $v_i$. The deformation $\delta$  and diameter of contact $a$ are generated by the collision with the viscoelastic board.}
\label{fig:F1}
\end{center}
\end{figure}

Here we think about the problems associated with a rigid sphere in free fall onto a viscoelastic board (See Fig.~\ref{fig:F1}). Under the experiment conditions previously explained, I assume that the adhesion effect is eliminated by dusting the board with chalk powder, thus only the viscoelastic bulk property contributes. In this case, the board is modeled by the viscoelastic-foundation model in which the stress deformation is described by foundations which are arranged in parallel\cite{Foundation}. The foundation model is a simplified model to describe the stress that is widely applied to the viscoelastic materials, appropriate when the half-space has a finite thickness. In my model, which I call Maxwell Viscoelastic Foundation model (MVF model), each foundation consists of a dashpot (viscous coefficient $\mu$) and a spring (elastic modulus $E$), which are serially connected. In this case, the stress $\sigma$ and the deformation $\epsilon$ can be related by the following differential equation with time $t$, $\frac{\mu}{E}\frac{d \sigma}{dt} + \sigma = \mu \frac{d \epsilon}{dt}$, which corresponds to the Maxwell model. By assuming that the deformation by the impact of the sphere is $\delta$ with a board thickness of board $h$, the deformation can be described by $\epsilon = \frac{\delta}{h}$. Thus, the rate of deformation is described by $\frac{d \epsilon}{dt} = \frac{1}{h}\frac{d \delta}{dt}$. In this model, it is assumed that the main contribution of deformation is due to $\frac{d \delta}{dt} \simeq {\rm const}$ for the foundation, which is supported by the experimental observations\cite{SMD}. Due to this condition, the differential equation is easily solved as $\sigma\left(t \right) = \mu \frac{d \delta}{dt}\left[1 - e^{-\frac{Et}{\mu}} \right]$. As the rate of deformation is independent of the position of contact within the contact area $\pi a^2$ where $a$ is the contact radius, thus the energy of deformation is described by 
\begin{equation}
E_{MVF} = \frac{\pi \mu \phi R \delta^2 }{h}\frac{d \delta}{dt} \left[ 1 - \exp \left(-\frac{E t_c}{ \mu} \right) \right] 
\label{eq:E1} 
\end{equation} 
where $R$ is radius of sphere, $\phi$ is the fraction of contact and $t_c$ is the contact time. $\phi = 1$ in the plane surface.

$E_{MVF}$ is quite characteristic as it transforms depending on the contact time $ \mu / E t_c =  {\rm De}$, Deborah number\cite{Reiner}. Supposing  ${\rm De} \gg 1$ which can be realized by the fast-time impact due to the following scaling $t_c \sim \delta_m / {v_i}$ where $\delta_m$ is maximum deformation, $v_i$ is the impact-velocity, and the relation of $\frac{ d \delta}{dt} = v_i$,  Taylor expansion is applied to $E_{MVF}$ as follows; $E_{MVF} = \frac{\pi \mu \phi R \delta_m^2 }{h} v_i \left[  \frac{E \delta_m}{\mu v_i} - \cdots \right] \simeq  \frac{\pi E \phi R \delta_m^3 }{h} = E_{el} $, which corresponds to the elastic energy\cite{Maruoka2019,Chastel2016}. Note that the scaling $t_c \sim \delta_m / v_i$ is easily deduced from the condition $\frac{d \delta}{dt} \simeq {\rm const}$. This result shows that $E_{MVF}$ experiences a transition to fully elastic energy or the energy mixed with viscous component depending on the contact time.

Suppose that the kinetic energy of the solid ball with the density $\rho$ is converted to $E_{MVF}$, we have 
\begin{equation}
\frac{2}{3}\pi R^3 \rho v_i^2  = \frac{\pi \mu \phi R \delta_m^2 }{h} v_i \left[ 1- \exp  \left( - \frac{ E \delta_m}{\mu v_i}\right) \right];
\label{eq:E2}
\end{equation}
which is the energy exchange at the maximum deformation.

\section{Analysis of self-similarity}

In this section, I intend to demonstrate how the MVF model gives rise to a crossover of scaling law through dimensionless analysis based on the framework proposed in the previous section. Eq.~(\ref{eq:E2}) suggests the possibility to change the form of energy depending on the contact time. Here, I intend to visualize the dynamics involved in this dynamical impact problem by exploring self-similar structures.

In order to see the self-similar structure, here I perform a dimensional analysis\cite{DimAna}. The physical parameters which are involved are summarized to the following function $ \delta_{m} = f \left(R, h, \phi, \rho, \mu, E, v_i \right)$. The dimensions of the function are described as follows; $[\delta_{m}] = L, [R] =L, [h] = L, [\phi] = 1, [\rho] = M/L^3, [\mu] = M/LT, [E] = M/LT^2, [v_i] = L/T$ by $LMT$ unit. By selecting $R, ~ \rho,~ E$ as the governing parameters with independent dimensions, which are defined as the parameters which cannot be represented as a product of the remaining parameters, the following similarity parameters are defined: 
\begin{equation}
\Pi = \frac{\delta_m}{R},~\kappa = \frac{h}{R}, ~\eta = \frac{\rho v_i^2}{E},~\theta = \frac{\mu}{E^{1/2} \rho^{1/2} R}
\label{eq:E3}
\end{equation}
then we have $\Pi = \Phi ( \phi, \kappa, \theta, \eta)$ where $\eta$ corresponds to the Cauchy number, a dimensionless velocity-component. 

Next, I proceed to the strategy proposed in the framework of crossover of scaling law in Fig. \ref{fig:F1a}. 

{\it ---Step 1: start from the scaling law valid in a certain scale region}. To go further to consider the self-similarity structure, the following solution is quite helpful, 
\begin{equation}
\Pi = {\rm const}\left(\frac{\kappa}{\phi} \right)^{\frac{1}{3}} \eta^{\frac{1}{3}}
\label{eq:E4a}
\end{equation}
which corresponds to Chastel-Gondret-Mongruel (CGM) solution\cite{Chastel2016,Chastel2019,Mongruel2020}. The CGM solution is obtained when the kinetic energy is fully transformed to elastic energy on a foundation model, which corresponds to the solution obtained from Eq.~(\ref{eq:E2}) in case of ${\rm De} \gg 1$. I have shown previously that $E_{MVF}$ turns to be the elastic energy in high-velocity impacts. It is expected that the scaling solution Eq.~(\ref{eq:E4a}) receives a kind of operation and transforms to another in low-velocity impacts in which ${\rm De} < 1$. This transformation is expected to give rise to a crossover. Following the framework mentioned in the previous section (Fig.~\ref{fig:F1a}), the CGR solution is appropriate for the scaling law in an idealized region. 

{\it  ---Step 2: define the dimensionless number composed of the scaling law}. Here we define newly a dimensionless number $\Psi = \frac{ \Pi^3 \phi }{\kappa \eta}$ from CGR solution (Eq.~(\ref{eq:E4a})). As it was discussed, $\Psi$ is constant when the CGR solution is valid, which holds true in the elastic region. However, $\Psi$ is not constant out of this region, and depends on another dimensionless number. 

 {\it ---Step 3: construct a self-similar solution by identifying an interfering dimensionless number}. Here, we identify the interfering physical quantity and construct the self-similar solution. The quantity that deviates it from an idealzed region is the viscous coefficient $\mu$ and its dimensionless number $\theta$. The dependence of $\Psi$ on $\theta$ can be identified from Eq.~(\ref{eq:E2}) by defining a new dimensionless number $Z = \frac{ \Pi}{ \theta \eta^{1/2}} =\frac{ E \delta_m}{ \mu  v_i}$ as follows,
\begin{equation}
\Psi =  \frac{2}{3}\frac{Z}{\left[ 1-\exp \left( -Z \right)\right] }
\label{eq:E4}
\end{equation}
though $Z$ equals to $1/{\rm De}$. Supposing $\Psi = \Phi \left(Z  \right)$, $\Phi$ converges to a finite limite as $Z$ goes to zero\cite{Conv}. Therefore, equation~(\ref{eq:E4}) belongs to a self-similar solution of the second kind\cite{SecondKind}, which is defined as the power-exponents of similarity parameters that cannot be determined by dimensional analysis and mathematically corresponds to a {\it fractal}\cite{Mandelbrot1983}. Here $\Psi = \Phi \left(Z\right)$ is the self-similar solution that describes the dependence of two dimensionless numbers and the crossover of scaling law in this problem.

\begin{figure}[h!]
  \includegraphics[width=1\columnwidth]{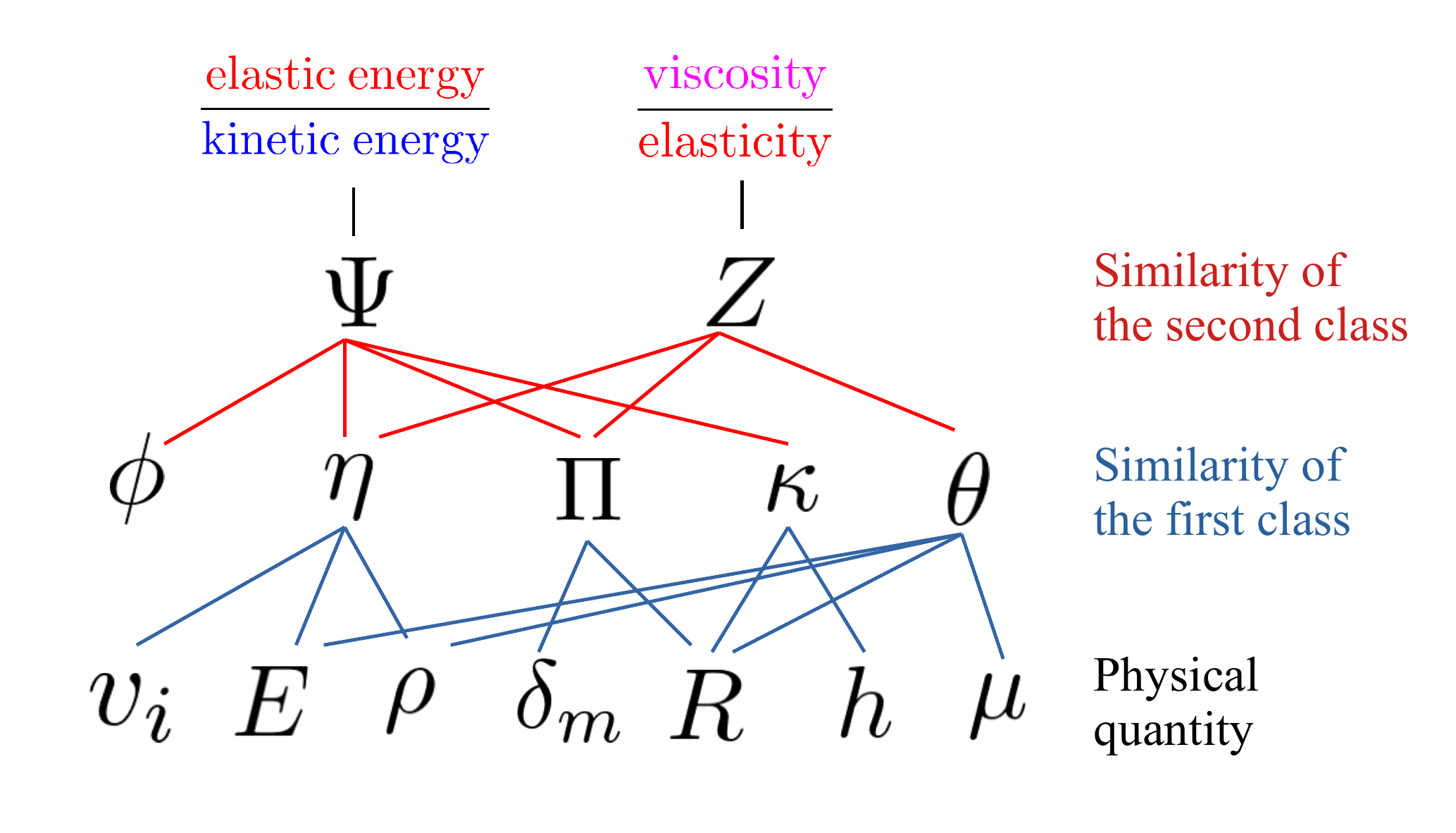}
  \caption{(Color online) The hierarchy of self-similarity on the dynamical impact of a solid sphere with a viscoelastic board. The solid lines signify the composition of each dimensionless parameter. In the present study, one finds that there is a hierarchy consisting of three classes of variables; the class to which physical quantities belong, the class to which dimensionless parameters composed by dimensional analysis belong and the class to which dimensionless parameters which is power-law monomial of dimensionless parameters to recover the convergence belong.  \label{fig:FS1b}}
\end{figure}

Note that there is a hierarchical structure on the self-similar solution in Eq.~(\ref{eq:E4}) depending on the convergence of dimensionless function (See Fig. \ref{fig:FS1b}). $\Pi$, $\kappa$, $\theta$, $\phi$ and $\eta$ belong to a similarity-class which forms the following similarity structure: $\Pi = \Phi ( \phi, \kappa, \theta, \eta)$ . Here I call a class similarity of {\it the first class} as it is generated through dimensional analysis. In the first class, each parameter belong to dimensionless physical quantities. On the other hand, $\Psi$ and $Z$ belong to another similarity-class to form the following similarity structure: $\Psi = \Phi \left( Z \right)$ where $\Psi = \frac{ \Pi^3 \phi }{\kappa \eta}$ and  $Z = \frac{ \Pi}{ \theta \eta^{1/2}}$. I call this class a similarity of {\it the second class}\cite{class}. The variables of the second class normalize the difference of the variables of the first class to integrate the single lines, which corresponds to the data collapse\cite{Bhattacharjee,Nakazato2018}. 

In the second class, similarity parameters represent the dynamics of energy involved in the process. $\Psi$ represents the proportion of kinetic energy and elastic energy while $Z$ represents the proportion of viscous energy and elastic energy, which corresponds to the reciprocal of the Deborah number. One can find that $\Phi(Z)$ represents the interference of viscous components. If $Z$ goes to 0, which can be achieved by the high-velocity impact or short-time contact, $\Phi \rightarrow {\rm const} $, then we have the CGM solution, which is realized in the case in which the kinetic energy fully transforms to elastic energy. Here the convergence of $\Phi(Z)$ means the inactiveness of $Z$. Thus in the case of $Z \ll 1$, the impact is elastically dominant, which we call elastic impact, giving 1/3 power-law on $\eta$. However, when this idealized condition is not satisfied ($Z >1$), which can be realized by ow-velocity impacts, the viscous components $\Phi(Z)$ interfere with $\Psi$. In this region, the viscosity contributes to the impact, and it changes the scaling law. This impact corresponds to a viscoelastic impact.

We cannot see the actual scaling behavior of $\Pi$ and $\eta$ in the viscoelastic regime from the second class. They belong to the first class and their behaviors are not simply consistent with $\Psi$ and $Z$ as one can see from $\frac{\Pi^3 \phi}{\kappa \eta} = \Phi \left( \frac{\Pi}{\theta \eta^{1/2}} \right)$, which shows that $\Pi$ and $\eta$ are included in both variables of the higher class, $\Psi$ and $Z$. In order to know the scaling-behavior of $\Pi$ and $\eta$, here I apply the perturbation method\cite{Holmes}, then we have
\begin{equation}
\Pi = \frac{\kappa}{54 \phi \theta^2 } + \left(\frac{\kappa^2 }{486 \phi^2 \theta^{3} }\right)^{\frac{1}{3}} \eta^{\frac{1}{6}} + \left( \frac{2 \kappa}{3 \phi} \right)^{\frac{1}{3}}\eta^{\frac{1}{3}}
\label{eq:E5}
\end{equation}
as $\varepsilon = \frac{1}{\theta \eta^{1/2}} \rightarrow 0$\cite{SMDer}.

Eq.~(\ref{eq:E5}) includes two different power exponents as $\eta^{1/3}$ and $\eta^{1/6}$, which suggests that intermediate asymptotics appear depending on $\theta$, $\eta$ or $Z$. In the case of the impact of high velocity and/or smaller sphere, which corresponds to $\eta \gg 1$ and/or $\theta \gg 1$ and $Z \ll 1$, $\eta^{1/3}$ is dominant. Conversely in the region of viscoelastic impact in which $Z > 1$, realized by a low-velocity impact $\eta \ll 1$ and/or the impact of larger sphere $\theta \ll 1$, $\eta^{1/6}$ is dominant while the intermediate behavior may be realized in $Z \sim 1$.  

\section{Result and discussion}
  
In the previous section, it was expected that the scaling law of the CGM solution experiences interference from the inverse Deborah number $Z$. This interference is described by Eq.~(\ref{eq:E4}) as a self-similar solution of the second kind, $\Psi = \Phi \left(Z \right)$. The self-similar solution directly describes the dynamics between the kinetic component, elasticity and viscosity. These dynamics are expressed in the similarity of the second class though actual scaling behavior is understood by Eq.~(\ref{eq:E5}) through the perturbation method, which predicts the existence of the crossover of scaling law between $\Pi$ and $\eta$. Each equation describes a different class of self-similarity.   

\begin{figure*}[t]
\begin{center}
\includegraphics[width=2\columnwidth]{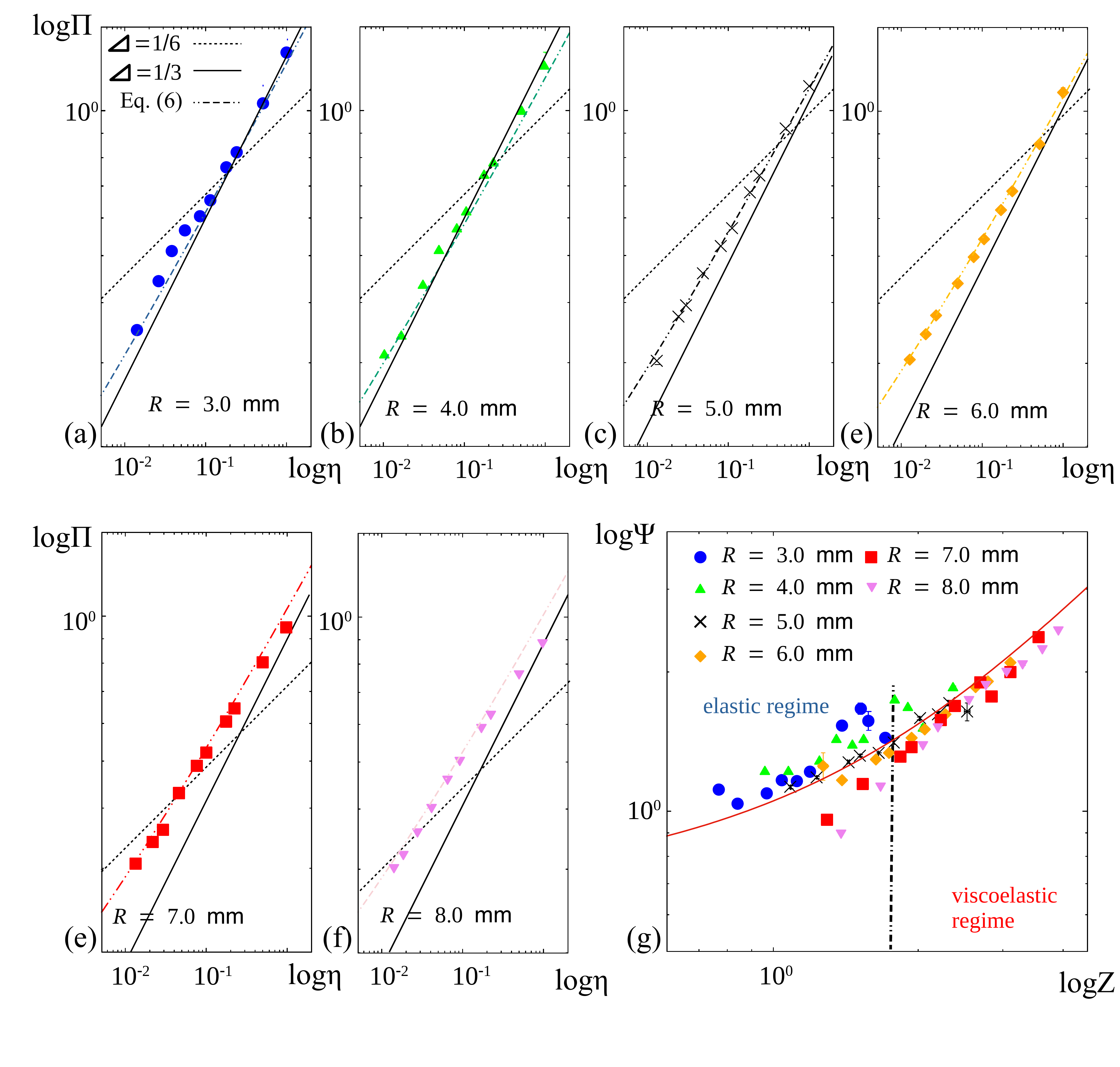}
\caption{(Color online) The different hierarchical structures of self-similarity. (a) - (f) Self-similarity of the first class : the power law relations $\Pi$ and $\eta$ for spheres of different sizes. The dashed lines indicate the slope of 1/6, the solid lines indicate the slope of 1/3 and the colored dot-dashed line indicates Eq.~(\ref{eq:E5}) for each size of the spheres. (g) Self-similarity of the second class : the plots between $\Psi$ and $Z$. $R = {\rm 3.0~mm}$ ($\textcolor{blue}{\bullet}$), ${\rm 4.0~mm}$ ($\textcolor{green}{\blacktriangle}$), {\rm 5.0~mm} ($\times$), {\rm 6.0~mm} ($\textcolor{orange}{\blacklozenge}$), {\rm 7.0~mm} ($\textcolor{red}{\blacksquare}$)  and {\rm 8.0~mm} ($\textcolor[rgb]{0.7, 0.4, 0.9}{\blacktriangledown}$) where $\Pi=\delta_m/R$, $\eta=\rho v_i^2/E$, $\Psi=\frac{\Pi^3 \phi}{\kappa \eta} = \frac{ \delta_m^3 E \phi}{ R^{2} h \rho v_i^2} $ and $Z = \frac{ \Pi}{ \theta \eta^{1/2}} = \frac{ E \delta_m }{ \mu  v_i}$. The red line in Figure (g) is Eq.~(\ref{eq:E4}). The dashed line roughly indicates the line separating the region. All the data points are calculated from the experimental results.}
\label{fig:F2}
\end{center}
\end{figure*}

Figure~\ref{fig:F2} shows the similarity parameters in different self-similarity classes. Figure~\ref{fig:F2} (a - f) demonstrates the self-similarity of the first class, which is the power-law relation between $\Pi$ and $\eta$ for different sizes of spheres while Figure~\ref{fig:F2} (g) demonstrates the self-similarity of the second class, which is their value for $\Psi$ and $Z$. As we can see, the plots of $\Pi$ and $\eta$ reveal a gradually different scaling law from $R=$ 3.0 mm to 8.0 mm. 

Equation (\ref{eq:E5}) predicts that $\Pi$ and $\eta$ have a different power law depending on $\theta$ and $\eta$, which is finally summarized to $\Phi(Z)$. As the prediction mentioned in the previous section suggests, the impacts of the smaller spheres ($R$ = 3.0, 4.0 mm) which tend to have smaller $Z$ follow the $1/3$ power-law, which corresponds to an elastic impact (Fig. \ref{fig:F2} (a, b)). The data points of the smaller spheres tend to converge to a finite limit, which signifies that the viscous component expressed by $\Phi(Z)$ hardly contributes. The data points of the intermediate size of spheres ($R$ = 5.0, 6.0 mm) reveal a deviation from the 1/3 power law, and the intermediate scaling laws between 1/3 and 1/6 though this behavior is well described by the Eq. (\ref{eq:E5}). These data points belong to the domain in which the viscosity contributes, as the contribution is expressed as the dependence of $Z$ in Fig.~\ref{fig:F2} (g). The larger spheres ($R$ = 7.0, 8.0 mm) reveal a larger deviation from the 1/3 power law of the elastic impact  (Fig. \ref{fig:F2} (e, f)). Their power laws are closer to 1/6 power-law, particularly in low-velocity. The data points revealing 1/6 power-laws belong to the larger $Z$ and larger $\Psi$ in Fig.~\ref{fig:F2} (g), which suggests the stronger viscous contribution. The dash-dots lines which are described by Eq.~(\ref{eq:E5}) are consistent with the power-law behavior of the data points. Figure ~\ref{fig:F2} (g) shows a good data collapse largely, and all the data points follow the theoretical line of Eq.~(\ref{eq:E4}).

We see that some data points deviate from the theoretical line for high-velocity impacts in the larger sphere ( $R >  7.0$ mm) with the small $Z$ in Fig.~\ref{fig:F2} (g). This may be because the indentation formed by high-velocity impacts of a larger sphere tends to be so intense that the assumption of the foundation model could be violated. In contact mechanics, the deformation is assumed to be smaller in comparison with the radius of the sphere. As the deviation from the theoretical line is larger in the case in which the indentation is large compared to the thickness of the board, the contact may be no longer elastic. However, other plots follow the single line well. We can say that the overall behaviors correspond well to the theory. The scaling behaviors were well described by theoretical lines in Fig.~\ref{fig:F2} (a-f), which demonstrates a good consistency in each size sphere.

The model assumes the main contribution of the rate of deformation is due to $\frac{d \delta}{dt} = v_i$, which corresponds to the square deformation. This assumption is well justified by observations (see Fig. S2-4)\cite{SMD}, which show that the rate of deformation corresponds to the impact velocity $v_i$,  and maintains the speed for a while, and then steeply decreases in the end (Fig. S2-4 (a)). The attractors of deformation are quite similar in different impact-velocity (Fig. S2-4 (b)). As the attractors of deformation are similar and $\frac{d \delta}{dt} \simeq v_i$, the contact time should be linear to the reciprocal of the impact-velocity. It was clearly observed in Fig. S5. This scaling relation makes the following dependence; the lower the impact-velocity is, the longer the contact time is. This dependence creates different features of impacts and different scaling laws. 

The adhesion effect plays an important role on the viscoelastic impact though in this work we focused on the role of the viscoelasticity derived from the bulk property. The adhesion effect is effectively eliminated by dusting the surface of the board with chalk powder\cite{Persson,RobThoms}. The viscoelastic contribution between the surface and the ball is smaller in the crack closing than in the crack opening\cite{Persson}. If the effect of adhesion had strongly remained, the data collapse would have been scattered though such a sign does not appear in Fig. \ref{fig:F2}. Considering these, it is expected that the adhesion effect is quite limited in this work. However, when the surface is coated with grease (WD-40) which is expected to decrease the adhesion effect, the ball is rebounded but the data collapse is scattered to some extent\cite{SurWD}. It suggested that dusting the surface is more effective than coating with grease to eliminate the adhesion effect. It was suggested that such a surface interaction in the dynamical impact is more clear in the process in which the ball takes off from the board though in this work, we have discussed the other half of the process of the dynamical contact, in which the ball contact the surface to the maximum deformation. The other side of processes with the entire process of dynamical impacts will be discussed in my future work.

For other interactions, Falcon et al. reported that the gravity effect on a repeated bouncing ball\cite{Falcon}. In this work, as the maximal deformation and the impact velocity, which is determined by the energy exchange, are focused on, the role of gravity is not apparent. However, it is indirectly related with the impact-velocity while the gravity field is constant in this experiment. This effect does not make difference on the results, which is the limitation of this work.

$E_{MVF}$ reveals an interesting feature as it transforms its form qualitatively depending on $Z$. It should be noted that such behavior does not appear from the Kelvin-Voigt model, which is another model for viscoelastic materials and consists of the spring and dash pot are parallelly connected. Therefore, the Kelvin-Voigt model does not create a crossover. It is clear that the Kelvin-Voigt model is not appropriate for this problem\cite{KVm}. On the other hand, it is found that the Zener model in which the Kelvin-Voigt model and the Maxwell model are combined, also reveals the similar crossover between $\eta^{1/3}$ and $\eta^{1/6}$ with a different coefficient on Eq.~(\ref{eq:E5}) by the perturbation method (Eq. (S28))\cite{Zm}. In the solution, a coefficient $K = \frac{3 \nu}{3 \nu +1}$ where $\nu = E/E_K$ is involved. $E_K$ is the elastic modulus derived from the parallelly connected spring. The Maxwell model is realized by $\nu \gg 1$, then the solution corresponds to Eq.~(\ref{eq:E5}). Kelvin-Voigt model can be realized by the opposite limit though $\nu \ll 1$ makes the solution vanish. It suggests that the Maxwell element is essential for the crossover. 

I proposed a framework for a crossover of scaling law; the interference of another dimensionless number generates a crossover (Fig. \ref{fig:F1a}). Note that the CGM solution is normalized to a dimensionless number $\Psi$, then $\Phi(Z)$ describes how $\Psi$ is interfered and changed by $Z$. This interference process is visualized in Fig.~\ref{fig:F2} (g) using the experimental results. In the end, the Maxwell model constitutes such interference as a self-similar solution Eq.~(\ref{eq:E4}). When $\Psi$ behaves as constant $\Phi = {\rm const}$,  the CGM solution is obtained as an intermediate asymptotics, then $\Pi \sim \eta^{1/3}$. However, in the region of $\Phi \neq {\rm const} $, which means the interference from another dimensionless number $Z$, another scaling law appears, $\Pi \sim \eta^{1/6}$. In this region, the viscoelastic effect starts to interfere. This is the fundamental mechanism of the crossover of scaling law. The self-similar solution,  Eq.~(\ref{eq:E4}) changes the CGM solution qualitatively. It is expected that such a class of self-similar solutions may exist on other crossovers of scaling law. Therefore, we expect; {\it there exists a self-similar solution on crossover of scaling law.} However, it should be noted that such a solution belongs to a higher class of the hierarchy of self-similarity as the dynamics involved in the problem is described by similarity parameters belonging to a higher class. It suggests that the hierarchy of self-similarity is quite important in a crossover of scaling law. 

We see that Eq.~(\ref{eq:E4}) describes the process of crossover continuously. Generally, in the studies of crossover, the scaling behaviors in each domain are investigated independently. However, in this work it was shown that there are three steps, starting from the idealized region in which a scaling law had already been obtained as the CGM solution, we then define a dimensionless number $\Psi$ from the CGM solution to identify the self-similar solution $\Phi$ from the MVF model. As we see previously, $\Phi$ describes the degree of transformation of scaling law which is normalized in $\Psi$. This strategy is quite unique and may be applicable to other problems. We generally can find an idealized region in which the problem is simpler and certain scaling law appears even though the non-idealized problem is difficult to attack. By starting from an idealized region and introducing the scaling law itself as a dimensionless number, one may find how this idealized region is changed, which may lead to a way to extend the problem into a non-idealized region. One notes that $\Phi$ does not only qualitatively decide the crossover but also it numerically expresses the balance between the dimensionless numbers. This numerical balance accurately decides the balance of coefficients of Eq. (\ref{eq:E5}), which enables us to describe the exact behaviors of crossover more accurately.

As the framework and mechanism of crossover are proposed, it is expected that such a framework may exist in other problems. All the stable scaling laws should be intermediate asymptotics in which dimensionless functions converge. Thus the transition of scaling laws must be given by the violation of this idealization. Yasuda et al. also reported the primal dimensional number to change the scaling laws\cite{Yasuda2020}. Barenblatt reported the dependence of power exponents by other dimensionless numbers\cite{Barenblatt2002,Barenblatt1981} though in these cases the exact form of the dimensionless function $\Phi$ was unclear and the hierarchy was not focused on. In the present work, I succeeded in identifying the exact form of dimensionless functions. The insights from the present work suggested that the investigation of the hierarchy of self-similarity can provide a clue to describe crossover.

Finally, the combination of dimensionless numbers listed in Eq.~(\ref{eq:E3}) is not the only possible selection. However, this combination is plausible in terms of perturbation. It suggests that the selection of dimensionless numbers is not arbitrary as it is related to the strategy of perturbation. $\theta = \frac{\mu}{E^{1/2} \rho^{1/2} R}$, appearing naturally in this problem and playing an essential role, is also an interesting dimensionless number as it can be expressed as $\theta = {\rm Re}/{\rm Ca}^{1/2}$ in which ${\rm Re}$ is the Reynolds number and ${\rm Ca}$ is the Cauchy number. $\theta$ used here indicates the proportion of viscosity, elasticity and inertia. 

\section{Conclusion}

The above discussion with experimental results confirms the validity of Eq.~(\ref{eq:E4}) with Eq.~(\ref{eq:E5}) as the fundamental equation of this problem. In this paper, I have successfully obtained the self-similar solutions governing the exact behavior of crossover of scaling law theoretically, which corresponds well to experimental results. Then I succeeded in demonstrating the framework and mechanism of the crossover of scaling law; {\it a crossover of scaling is generated by the interference of other similarity parameters of higher class}, which corresponds to the framework mentioned in the introduction. It suggests that there always exists a self-similar solution to crossover.

The method exercised in this work is unique in terms of methodology. This work successfully described the crossover of scaling law as a continuous process. The degree of this interference is quantitatively and qualitatively estimated and it enables us to describe crossover more accurately. This accuracy was guaranteed by the coefficients of the different scaling laws which were given by the self-similar solution of the second class, which suggests that the coefficients are essential for the accurate description of crossover. 

Finally, intermediate asymptotics is the {\it locally} valid asymptotic expression while we have found that this locality is governed by the self-similar solution of the higher class in this paper. This framework is simple and expected to be quite general in physics. Besides, it is similar to critical phenomena in which the transition of phase is generated by the continuous parameter variation. Therefore, the present work supplies interesting insights into the concept of self-similarity, nonequlibrium theory and critical phenomena, for a wide variety of fields in physics in general. 

\backmatter

\bmhead{Supplementary information}

\bmhead{Data availability}

The datasets generated during and/or analyzed during the current study are available from the corresponding author on reasonable request.

\bmhead{Acknowledgments}

The author wishes to thank D. Nishiura for technical assistance of the experiments, G. Li for technical advice of experimental setup, D. Matsuoka for technical advice for the algorithm of the program for image analysis, A. D. Sproson for advise for elaboration of the mauscript,  T. Yamaguchi, S. Okada, E. Barbieri and Y. Kawamura for helpful discussion and many support for the experiment. The author appreciates the members of the MAT (Center for Mathematical Science and Advanced Technology) seminar at JAMSTEC for their fruitful discussion.








\end{document}


\title{Supplemental materials to the manuscript of A framework of crossover of scaling law : dynamical impact of viscoelastic surface}
\maketitle

\section{Intermediate asymptotics}

\begin{figure}[h!]
\includegraphics[width=8.6cm]{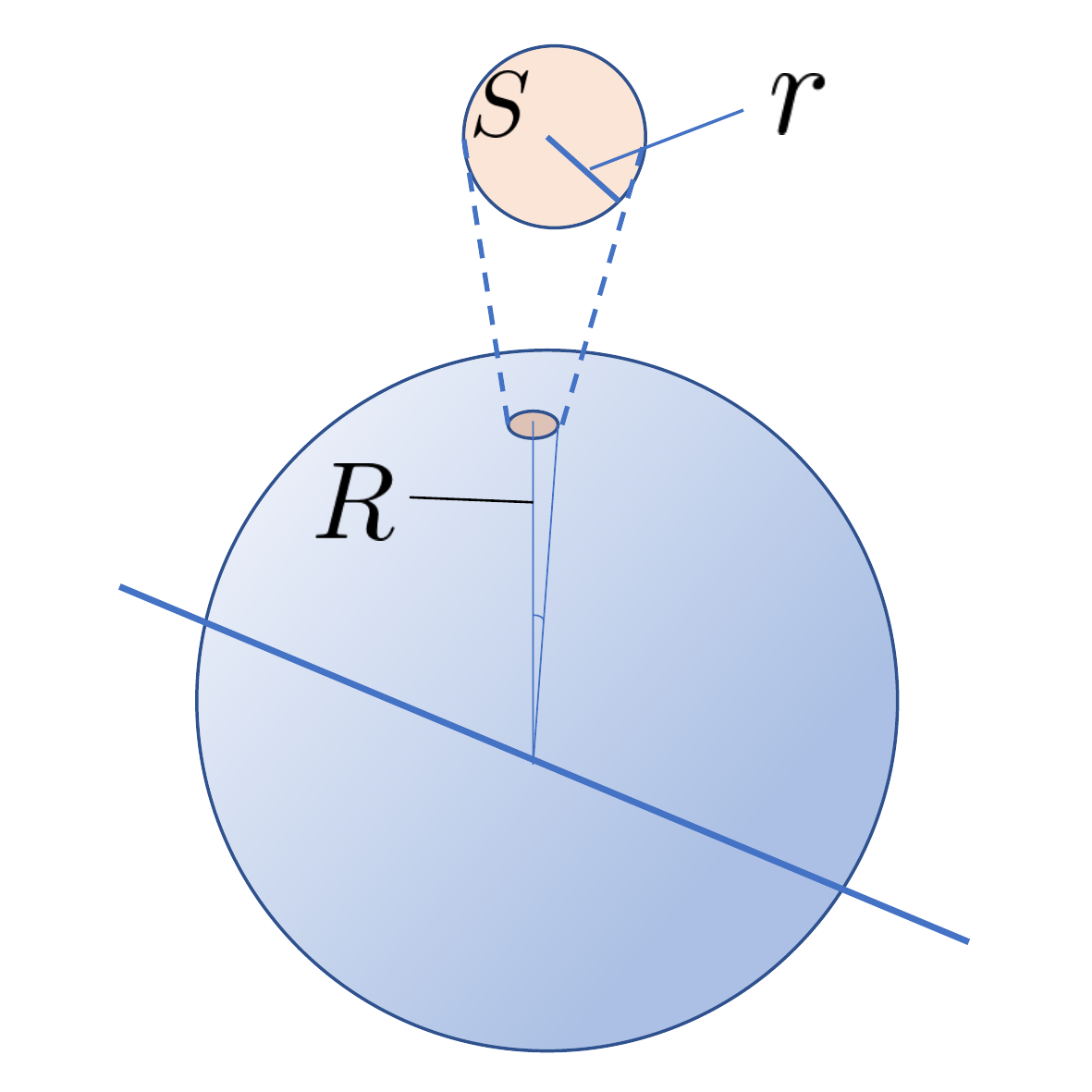}
\caption{(Color online) A circle of which radius is $r$ and surface area is $S$, described on a sphere of which radius is $R$.  \label{fig:FS1a}  }
\end{figure}

In this section, I briefly explain the concept of {\it intermediate asymptoitcs} which is formalized by Barenblatt\cite{Barenblatt2003,Barenblatt1996, Barenblatt2014,Barenblatt1972} by using a simple example. Intermediate asymptotics is an asymptotic representation of a function valid in a certain range of independent variables, which corresponds to a kind of the formalization of the idealization which accompanies with the construction of physical model. To understand this concept, the following problem of dimensional analysis might be helpful. Imagine that the circle is pictured on the surface of the sphere (See Fig.~\ref{fig:FS1a}). In this problem, the physical parameters that are involved are the surface area of circle $S$, radius of the circle $r$ and the radius of sphere $R$. Here we would like to know the scaling behavior between $S$ and $r$. Therefore we assume the functional relation as follows: $S = f(r, R)$. 

In this case, we attempt to obtain the exact scaling behavior by dimensional consideration. According to dimensional analysis, as the dimension of physical parameters $[S]=L^2$, $[r]=L$ and $[R]=L$, selecting $r$ as a governing parameter of independent dimension, we have the following dimensionless function,
\begin{equation}
\Pi = \Phi \left( \theta \right)
\label{eq:ES1}
\end{equation}
where $\Pi = \frac{S}{r^2}$ and $\theta = \frac{r}{R}$. Eq.~\ref{eq:ES1} suggests that we expect the following scaling relation, $S \sim r^2$, if $\Pi$ is constant. However, we easily find that this guess depends on the behavior of $\Phi$. 

By the geometrical consideration, in this case, we can calculate the exact form of $\Phi$ as follows,
\begin{equation}
\Phi\left(\theta\right) =2\pi \frac{1- {\rm cos} \theta}{\theta^2}.
\label{eq:ES2}
\end{equation}
To know the behavior of $\Phi$ in the case in which $\theta \rightarrow 0$, which corresponds to the increase of $R$ or the decrease of $r$, Taylor expansion is applied to Eq.~(\ref{eq:ES2}) then we have,
\begin{equation}
\Pi =\Phi\left(\theta\right) \simeq \pi -\frac{\pi}{12}\theta^2 \cdots +\underset{\theta \rightarrow 0}{\longrightarrow} \pi.
\label{eq:ES3}
\end{equation}
As Eq.~(\ref{eq:ES3}) shows, $\Phi$ converges to a finite limit $\pi$, then we have a following intermeidate asymptotics as $\Pi = \frac{S}{r^2}$,
\begin{equation}
S = \pi r^2~\left(0 < r \ll R\right)
\label{eq:ES4}
\end{equation}
as far as the asymptotic condition $\theta \ll 1$, corresponding to $0 < r \ll R$, is satisfied.

Note that the scaling law Eq.~(\ref{eq:ES4}) is valid in the scale range  $(0 < r \ll R)$, in which the circle is significantly smaller than the sphere. Therefore, Eq.~(\ref{eq:ES4}) is an asymptotic expression which is valid in the certain range of variable $r$. This scaling law formalized {\it locally} is an intermediate asymptotic in this problem. Barenblatt insisted that this framework is applicable to the construction of a physical model.

The important point of this concept is that every physical problem has dimension and can be applied to dimensional analysis to obtain dimensionless function $\Phi$. By considering the convergence of $\Phi$, some similarity parameters can be selected to have the idealized solution effectively and practically as the convergence of $\Phi$ can be verified by the experimental or simulational results even if the exact form of $\Phi$ is not obtained. This procedure gives rise to the strategy of Barenblatt as it is formalized in the recipe\cite{Recipe}.

The second important point is that this process, in which one screens the similarity parameters of $\Phi$ depending on their convergence, corresponds to an idealization of problems. More or less, all the physical models involve idealizations such as ignorance of friction force, ignorance of quantum or relativity effect. All these assumptions corresponds to the idealizing process of dimensionless function. For example, the ideal gas equation can be considered as an intermediate asymptotic valid in the range where the volume of molecules $b$ and the molecular interaction $a$ are negligible on van der Waals equation as follows,
\begin{equation}
p=\frac{nRT}{V-nb}-\frac{an^2}{V^2} \longrightarrow \frac{nRT}{V}~~\left(\frac{an^2}{V^2} \ll p \ll \frac{RT}{b} \right).
\label{eq:ES5}
\end{equation}
This idealizing scale range is satisfied as far as $\Pi_a = \frac{a n^2}{pV^2} \ll 1$ and $\Pi_b = \frac{pb}{RT} \ll 1$. The interested readers may refer to Ref.\cite{Oono2013} for further discussion related with phenomenology, renormalization and asymptotic analysis on physics.

This concept suggests that every physical theory is {\it locally} valid. This localization is quantitatively and qualitatively formalized by the intermediate asymptotics. In the present work, the author focuses on this point and considers the case of the transition of this {\it locality}.

\section{Complete similarity and incomplete similarity}

In this section, I briefly explain {\it complete similarity} and {\it incomplete similarity}, as well as the {\it self-similarity of the first kind} and the {\it self-similarity of the second kind} \cite{SecondKind}. They are the category in terms of the convergence of dimensionless function. Zeldovich noted that there exists the type of self-similarity\cite{Zeldovich1956}. As the previous section showed, the self-similar solution is obtained by dimensional analysis. Supposing that a certain physical function, 
\begin{equation}
y = f(t,x,z)
\label{eq:ES6}
\end{equation}
in which $y$, $t$, $x$ and $z$ are certain physical quantities which have physical dimensions. Selecting $t$ as a governing parameter with independent dimension, which is defined as physical parameters which cannot be represented as a product of the remaining parameters, then we apply dimensional analysis to have
\begin{equation}
\Pi = \Phi(\eta, \xi)
\label{eq:ES7}
\end{equation}
where $\Pi = y/t^{\alpha}$, $\eta = x/t^{\beta}$ and $\xi = z/t^{\gamma}$. $\alpha$, $\beta$ and $\gamma$ can be fully determined by the consideration of dimension of parameters through dimensional analysis.

If $\Phi$ converges to a finite limit as $\xi$ goes to zero or infinity, this case corresponds to {\it complete similarity} or {\it similarity of the first kind} in the similarity parameter $\xi$. In this case, $\xi$ can be excluded on the consideration and we have an intermediate asymptotics. Once $\eta$ and $\xi$ both satisfy the complete similarity then $\Phi \rightarrow {\rm const}$ as $\eta \gg 1$ and $\xi \gg 1$, then we have a following intermediate asymptotic, $y = {\rm const}~t^{\alpha}~(0 < t \ll x^{1/\beta},~0 < t \ll z^{1/\gamma})$. When the similarity parameters satisfies the condition of complete similarity, $\Pi = \Phi( \xi, \eta )$ is corresponds to a {\it self-similar solution of the first kind}. In the previous section, as Eq.~(\ref{eq:ES3}) shows, the dimensionless function converges to a finite limit, therefore the problem belongs to complete similarity and Eq.~(\ref{eq:ES1}) is a self-similar solution of the first kind.

On the other hand, in the case in which the complete similarity is not satisfied, namely $\Phi$ does not converge to a finite limit as $\eta$ goes to zero or infinity, but the convergence is recovered by constructing new similarity parameters as the power-law monomial using dimensionless variables, this case corresponds to {\it incomplete similarity} or {\it similarity of the second kind}. In this case, we may have the following self-similar solution, which is called {\it self-similar solution of the second kind},
\begin{equation}
\Psi = \Phi(Z)
\label{eq:ES8}
\end{equation}
where $\Psi = \Pi / \eta^{\zeta}$ and $Z = \xi / \eta^{\epsilon}$. 

The first important point is that the power exponent $\zeta$ and $\epsilon$ cannot be determined by dimensional analysis in the case of the second kind while it is possible in the case of the first kind. We may occasionally determine $\zeta$ and $\epsilon$ by the method for nonlinear eigenvalue problems\cite{NEP} or renormalization group theory\cite{Goldenfeld1992,Goldenfeld} though we may consider them as simply empirical numbers\cite{Barenblatt1981}. It was suggested that self-similarity of the second kind corresponds to {\it fractal}\cite{fractal, Mandelbrot1983}, which was elaborated by Mandelbrot. The fractal is a geometrical object which is lacking in a characteristic length. If the objects possess a certain characteristic length, the scale of the object is apparent by scale transformation. On the other hand, the scale of fractal objects is not apparent but self-similar as the scale transformation. Such a geometrical property corresponds to the divergence of dimensionless function in dimensional analysis. 
 
The second important point is that there exists a hierarchy of self-similarity. Note that we can find a parallelism between the first kind and the second kind. Dimensional analysis transforms $y=f(t,x,z)$ to $y/t^{\alpha} = \Phi(x/t^{\beta}, z/t^{\gamma})$ while $\Pi = \Phi(\eta, \xi)$ is transformed to  $\Pi / \eta^{\zeta} = \Phi( \xi / \eta^{\epsilon} )$ in case of the second kind. In the present study, I focused on this hierarchy though the first kind and the second kind refer to the property of the convergence of dimensionless function, not to the classes to which dimensionless parameters belong. Thus, I introduced a word, {\it class} to characterize the hierarchical structure.

By considering the convergence of the similarity parameters,  the self-similar structure of the problem are explored, and intermediate asymptotics is finally obtained. Depending on the type of similarity, the asymptotics is called {\it intermediate asymptotics of the first kind} or {\it intermediate asymptotics of the second kind}.

\section{The time evolution of deformation}

In this work, the model assumes the main contribution of deformation is due to $\frac{d \delta}{dt} = v_i$, which corresponds to the square deformation. This behavior is well supported by the observation of experiment. Fig.~S2-4 (a) shows the time evolution of deformation for different sizes of spheres, $R$ = 4.0 mm (Fig. S2), $R $ = 6.0 mm (Fig.~S3), $R$ = 8.0 mm (Fig.~S4). After the contact, the rate of deformation corresponds to the impact velocity and the rate of deformation is maintained for a while then it steeply decreases in the end. Fig.~S2-4 (b) shows the comparison of each impact for normalized deformation and time. The deformation is normalized by the maximum deformation $\delta_m$ and the time is normalized by $t_{{\rm max}}$ at which $\delta$ reaches $\delta_m$. One can find that all the attractors overlap almost completely, which signifies the attractors in different impact speeds are similar. This means that the lower the impact-velocity is, the longer the contact time is. This relation is well observed in Fig. ~\ref{fig:Fig2C}, which shows the linear scaling relation between $t_{{\rm max}}$ and $\delta_m /v_i$ in different sizes of spheres. The scaling relation between $t_{\rm max}$ and $v_i$ is strong relation valid in every data in different sizes of spheres whether they belong to the elastic regime or viscoelastic regime. This scaling relation makes the velocity-contact time dependence which generates the crossover of scaling law.
\begin{figure}[h!]
\subfigure[The time evolution of deformation for $R = 4.0 ~{\rm mm}$]{
 \includegraphics[width=7.5cm]{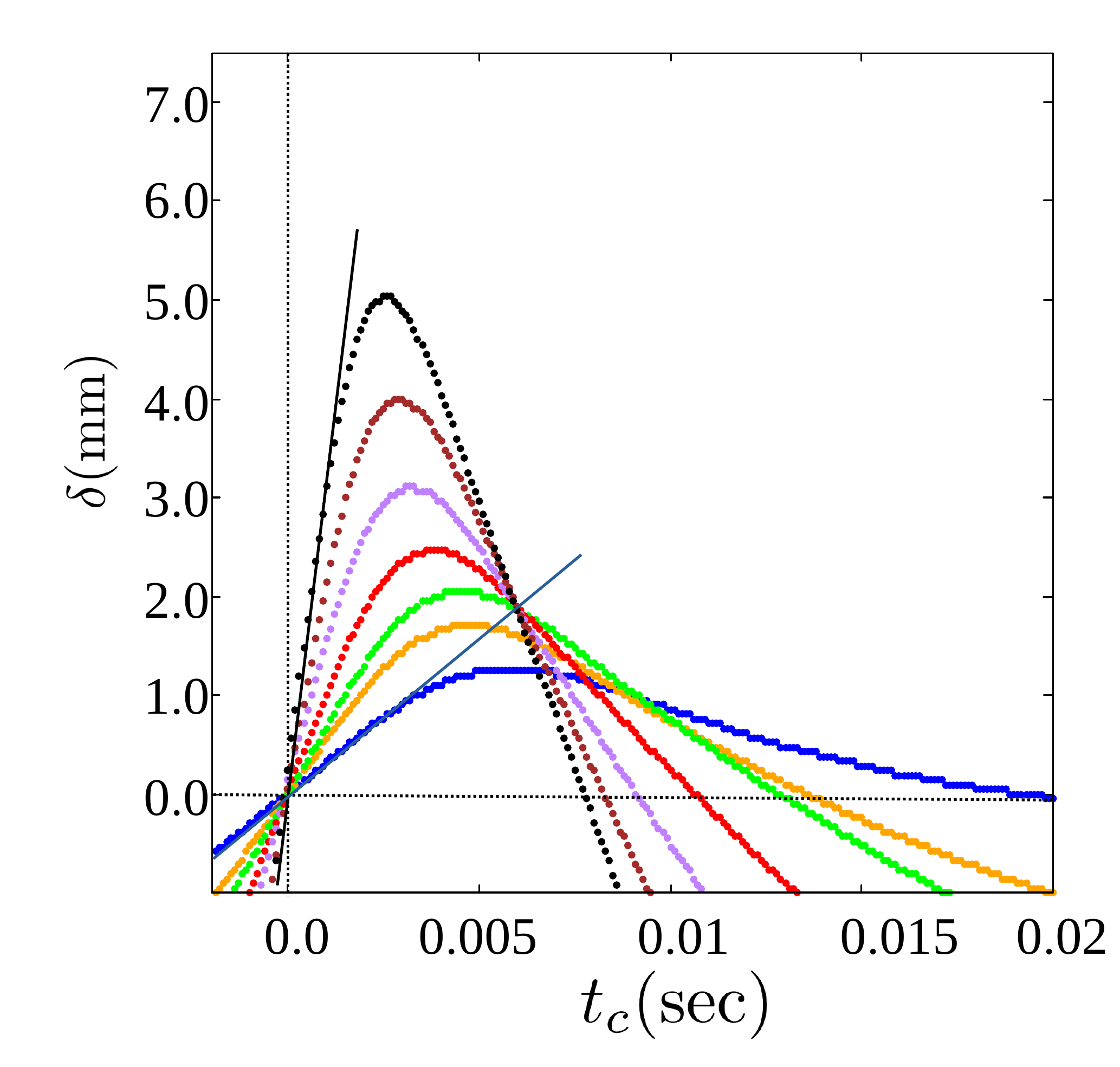}
\label{fig:Fig2A}}
\subfigure[The time evolution of normalized deformation for $R = 4.0 ~{\rm mm}$ ]{
  \includegraphics[width=7.5cm]{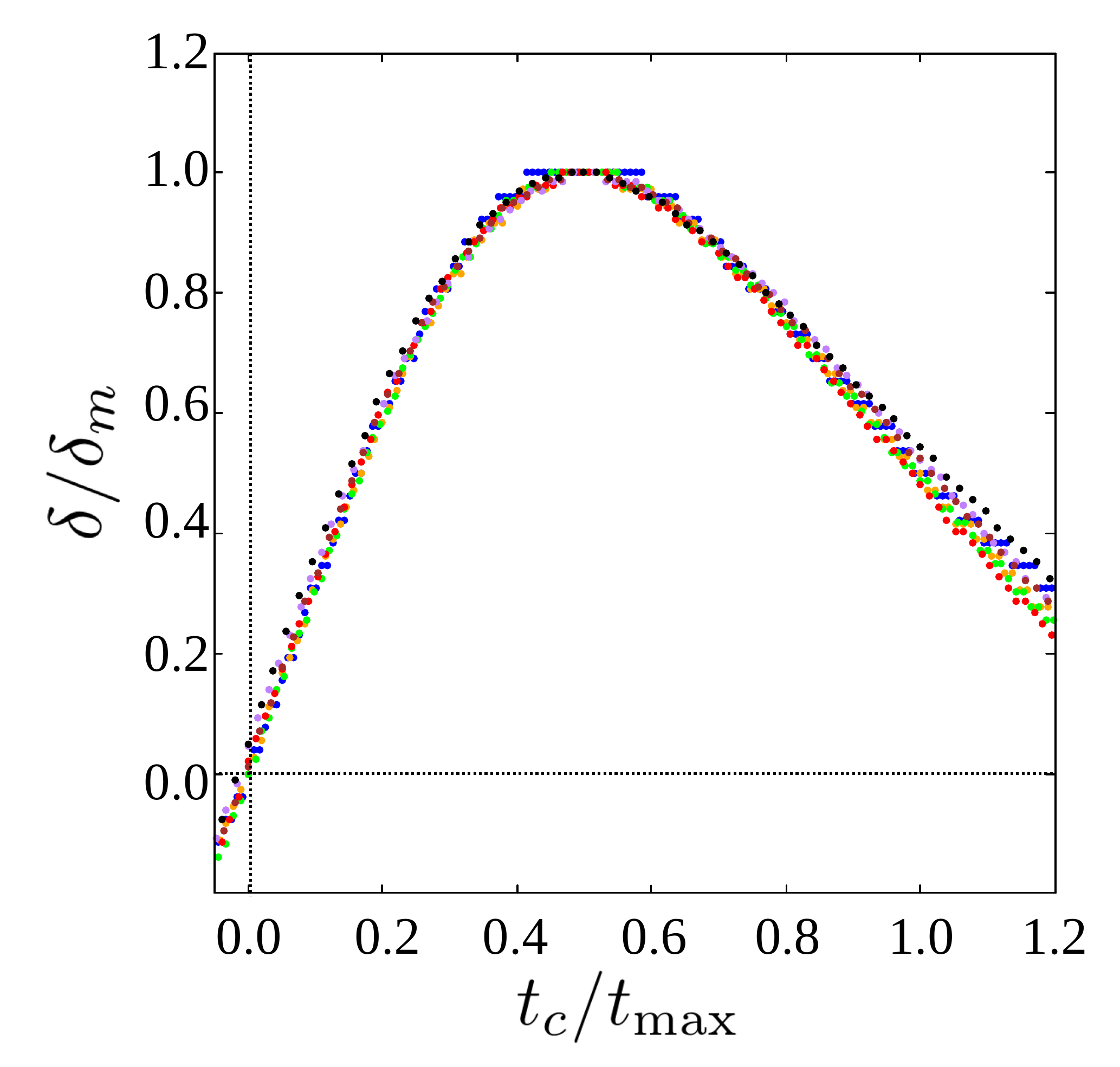}
\label{fig:Fig2B}}
 \caption{(Color online) (a) The time evolution of deformation $\delta$ and (b) the normalized deformation $\delta/ \delta_m$ for $R = 4.0 ~{\rm mm}$, $v_i = 316~{\rm mm/s}$, $Z = 2.38$ ($\textcolor{blue}{\bullet}$),  $v_i = 549~{\rm mm/s}$, $Z = 1.89$  ($\textcolor{orange}{\bullet}$), $v_i = 692~{\rm mm/s}$, $Z = 1.79$ ($\textcolor{green}{\bullet}$), $v_i =1021 ~{\rm mm/s}$, $Z = 1.46$ ($\textcolor{red}{\bullet}$), $v_i =1510~{\rm mm/s}$, $Z =1.25 $ ($\textcolor{violet}{\bullet}$), $v_i =2238~{\rm mm/s}$, $Z =1.25 $ ($\textcolor{brown}{\bullet}$) and $v_i = 3110~{\rm mm/s}$, $Z =0.98$ ($\textcolor{black}{\bullet}$). $\delta_m$ is maximum deformation and $t_{ {\rm max}}$ is the time at which $\delta$ reaches to $\delta_m$. The vertical dashed line indicates the moment of contact time and the horizontal dashed line indicates the configuration at $\delta = 0$. \label{fig:FS8} }
\end{figure}
\begin{figure}[h!]
\subfigure[The time evolution of deformation for $R = 6.0 ~{\rm mm}$]{
 \includegraphics[width=7.5cm]{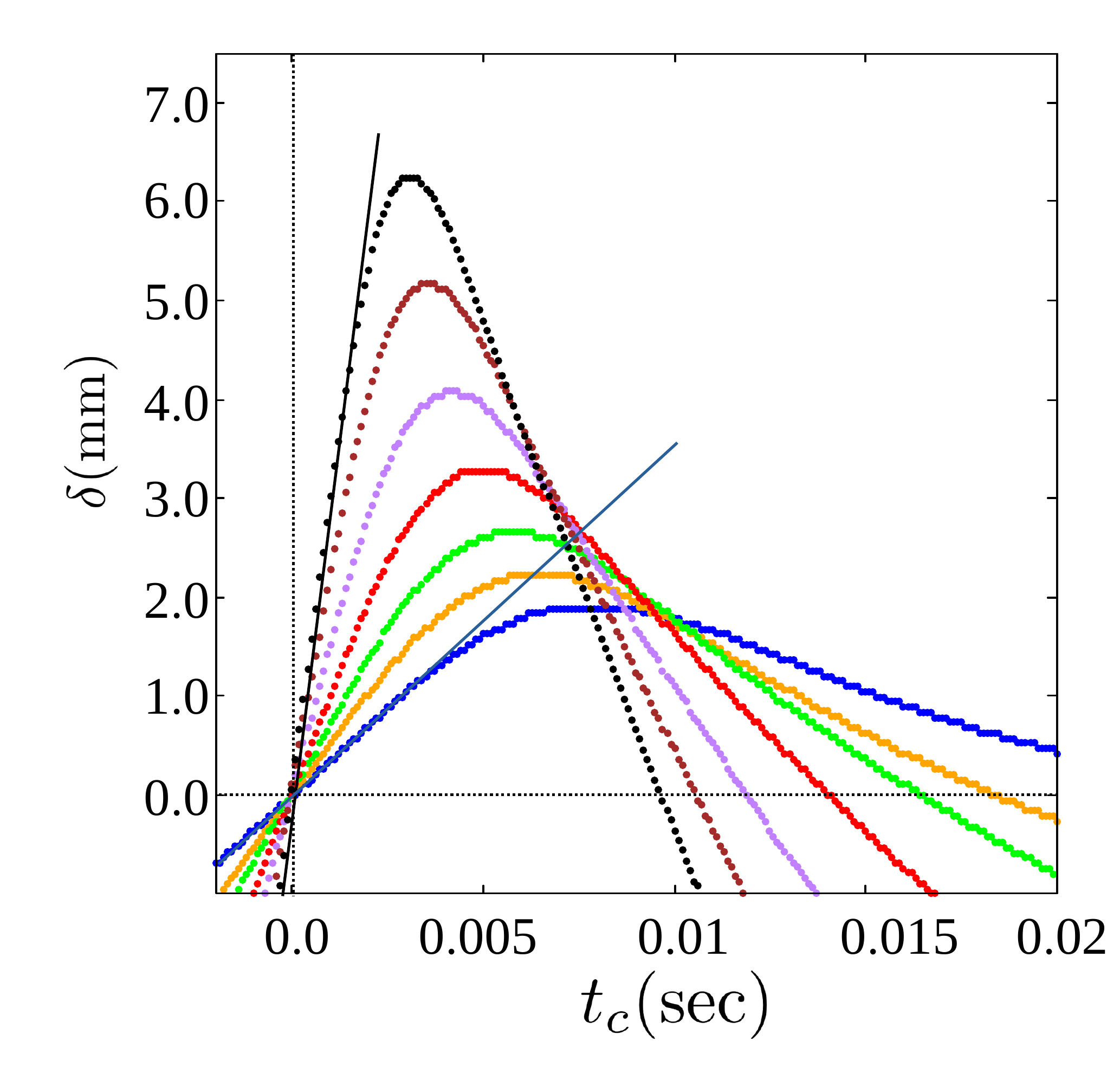}
\label{fig:Fig2A_12}}
\subfigure[The time evolution of normalized deformation for $R = 6.0 ~{\rm mm}$ ]{
  \includegraphics[width=7.5cm]{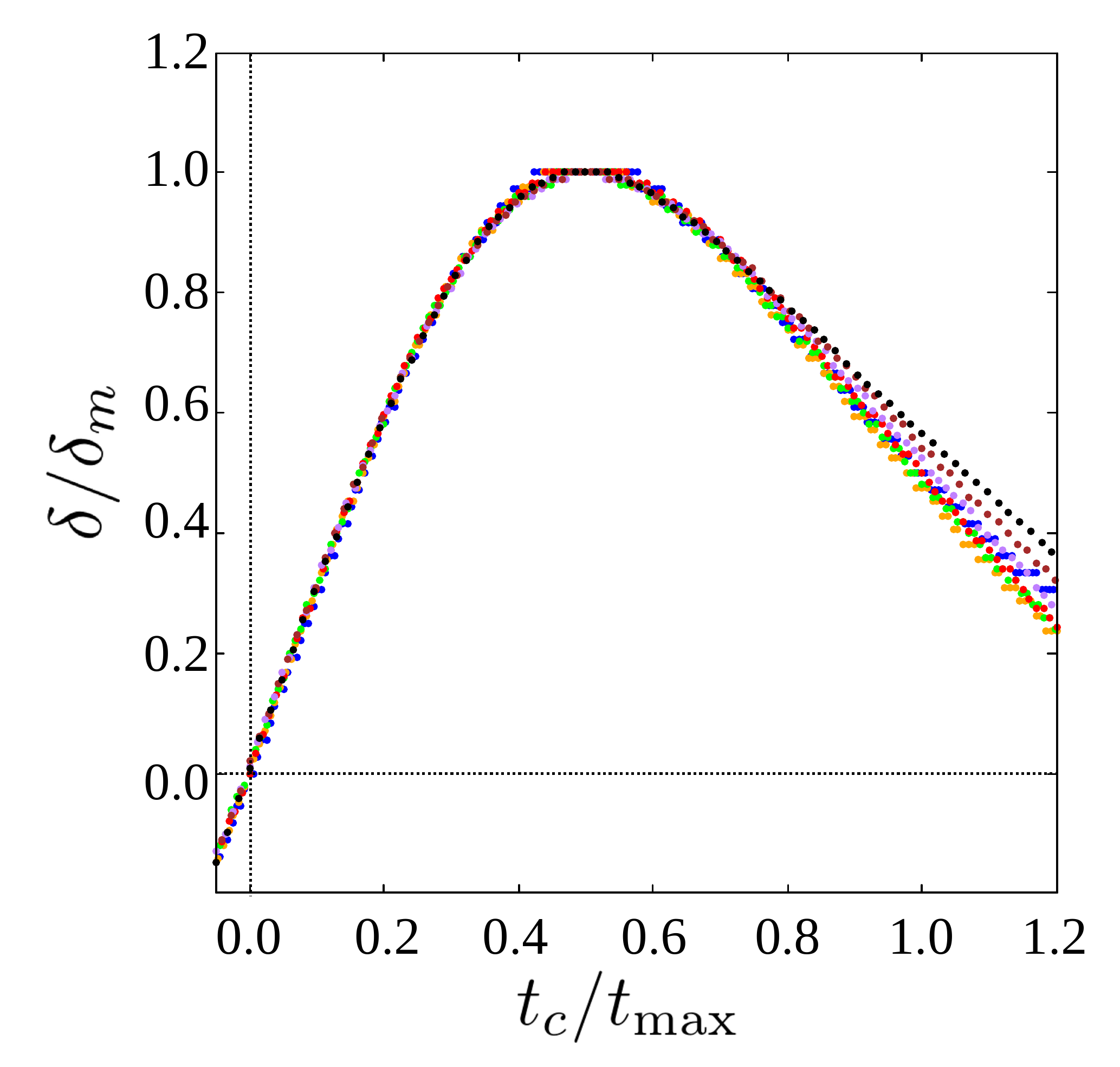}
\label{fig:Fig2B_12}}
 \caption{(Color online) (a) The time evolution of deformation $\delta$ and (b) the normalized deformation $\delta/ \delta_m$ for $R = 6.0 ~{\rm mm}$, $v_i = 351~{\rm mm/s}$, $Z = 3.24$ ($\textcolor{blue}{\bullet}$),  $v_i = 519~{\rm mm/s}$, $Z = 2.58$  ($\textcolor{orange}{\bullet}$), $v_i = 697~{\rm mm/s}$, $Z = 2.30$ ($\textcolor{green}{\bullet}$), $v_i =1013 ~{\rm mm/s}$, $Z = 1.95$ ($\textcolor{red}{\bullet}$), $v_i =1514 ~{\rm mm/s}$, $Z =1.63 $ ($\textcolor{violet}{\bullet}$), $v_i =2240~{\rm mm/s}$, $Z =1.40$ ($\textcolor{brown}{\bullet}$) and $v_i = 3125~{\rm mm/s}$, $Z =1.20$ ($\textcolor{black}{\bullet}$). $\delta_m$ is maximum deformation and $t_{ {\rm max}}$ is the time at which $\delta$ reaches to $\delta_m$. The vertical dashed line indicates the moment of contact time and the horizontal dashed line indicates the configuration at $\delta = 0$. \label{fig:FS8} }
\end{figure}
\begin{figure}[h!]
\subfigure[The time evolution of deformation for $R = 8.0 ~{\rm mm}$]{
 \includegraphics[width=7.5cm]{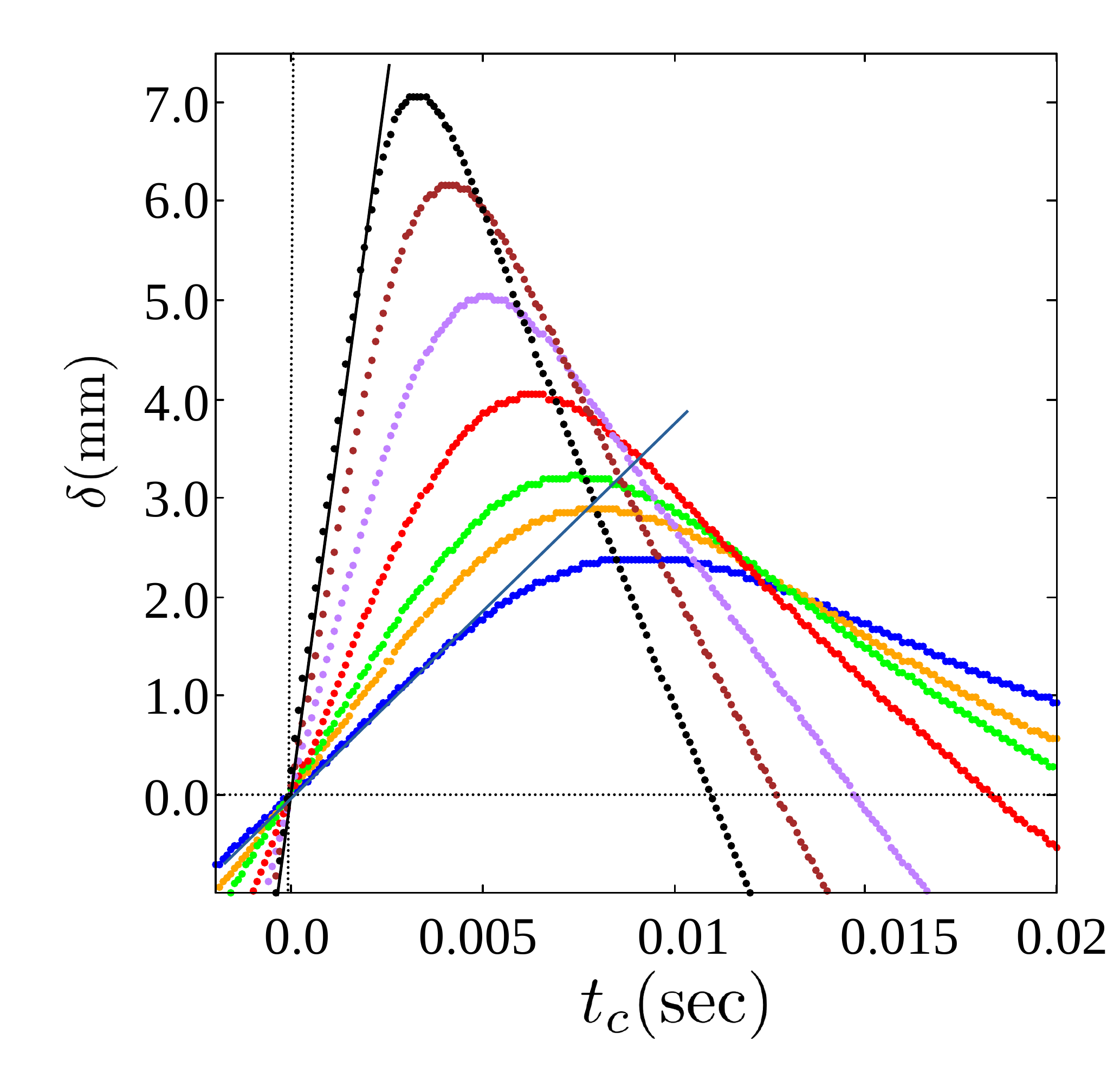}
\label{fig:Fig2A_16}}
\subfigure[The time evolution of normalized deformation for $R = 8.0 ~{\rm mm}$ ]{
  \includegraphics[width=7.5cm]{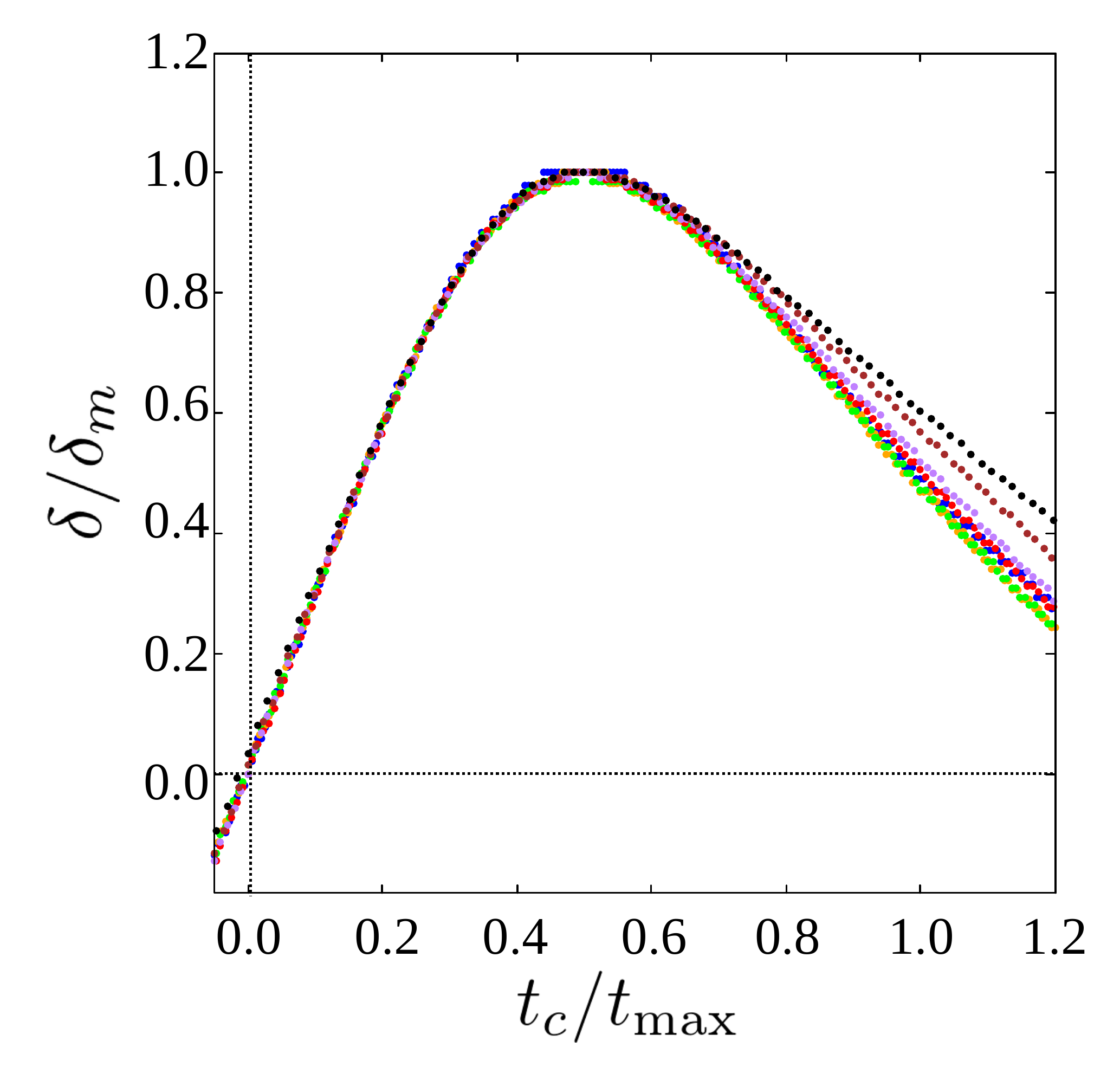}
\label{fig:Fig2B_16}}
 \caption{(Color online) (a) The time evolution of deformation $\delta$ and (b) the normalized deformation $\delta/ \delta_m$ for $R = 8.0 ~{\rm mm}$, $v_i = 372~{\rm mm/s}$, $Z = 3.86$ ($\textcolor{blue}{\bullet}$),  $v_i = 524~{\rm mm/s}$, $Z = 3.34$  ($\textcolor{orange}{\bullet}$), $v_i = 637~{\rm mm/s}$, $Z = 3.07$ ($\textcolor{green}{\bullet}$), $v_i =959 ~{\rm mm/s}$, $Z = 2.55$ ($\textcolor{red}{\bullet}$), $v_i =1492 ~{\rm mm/s}$, $Z =2.04 $ ($\textcolor{violet}{\bullet}$), $v_i =2219~{\rm mm/s}$, $Z =1.68 $ ($\textcolor{brown}{\bullet}$) and $v_i = 3104~{\rm mm/s}$, $Z =1.37$ ($\textcolor{black}{\bullet}$). $\delta_m$ is maximum deformation and $t_{ {\rm max}}$ is the time at which $\delta$ reaches to $\delta_m$. The vertical dashed line indicates the moment of contact time and the horizontal dashed line indicates the configuration at $\delta = 0$. \label{fig:FS8} }
\end{figure}
\begin{figure}[h!]
\includegraphics[width=8.6cm]{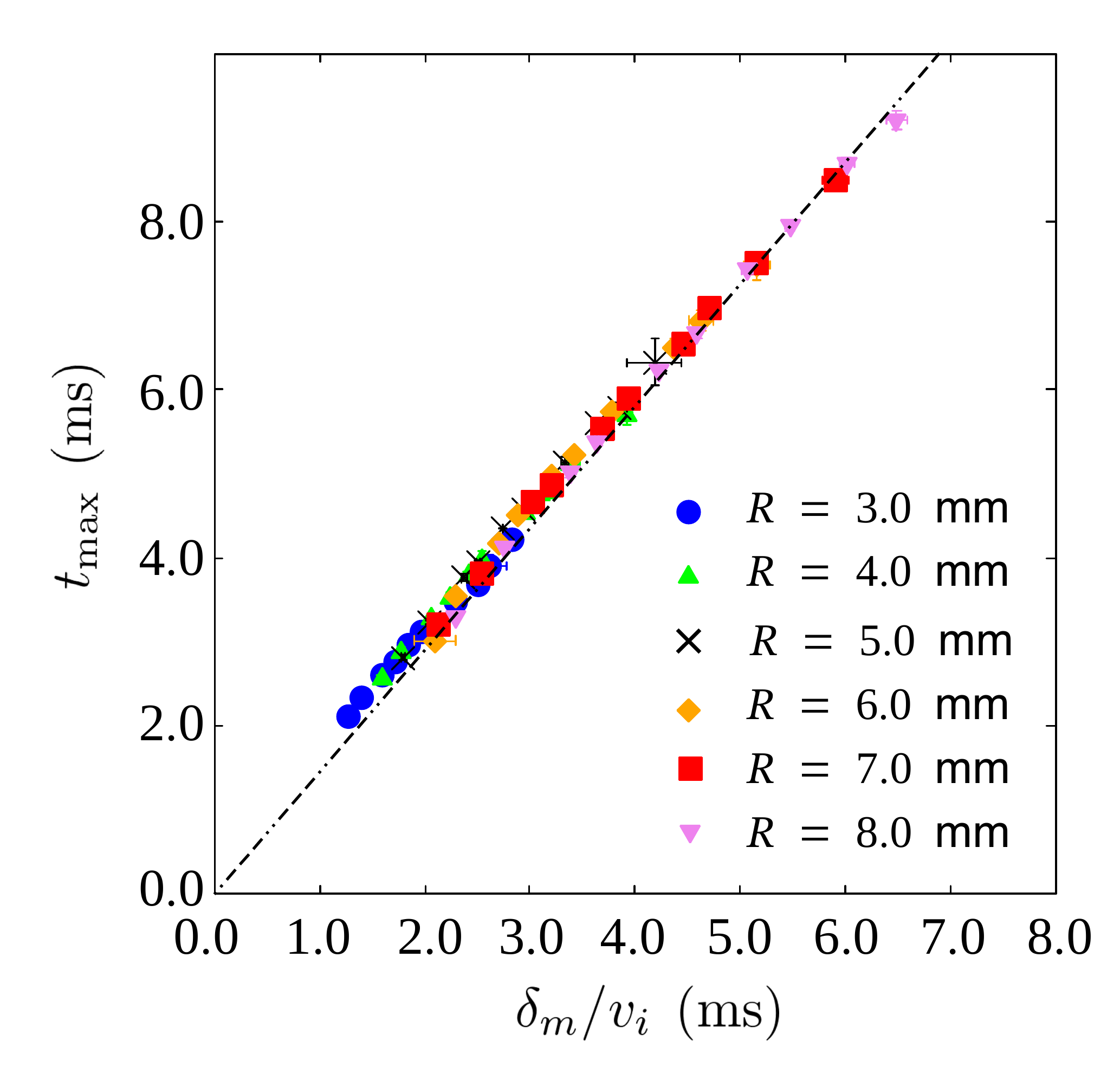}
\caption{(Color online) The dependence between $t_{ {\rm max}}$ and $\delta_m / v_i$ for each size of sphere.  \label{fig:Fig2C}  }
\end{figure}

\section{The convergence of Eq. (6).}

Eq. (6) is seemingly an indeterminate form as $Z \rightarrow 0$ though it converges to a finite limit as follows. Using L'H{\^{o}}pital's rule, then we have
\begin{equation}
\lim_{Z \to 0} \frac{2}{3} \frac{Z}{1-e^{-Z}} = \lim_{Z \to 0} \frac{2}{3} \frac{\left(Z \right)^{'}}{\left(1-e^{-Z}\right)^{'}} =  \frac{2}{3} .
\label{eq:ES8a}
\end{equation}

\section{The derivation of Eq. (7)}

According to Eq.~(6), in the self-similarity of the first class, we have
\begin{equation}
\frac{\Pi^3 \phi}{\kappa \eta} = \frac{2}{3}\frac{ \frac{\Pi}{\theta \eta^{1/2}} }{\left[1 - \exp \left( -\frac{\Pi}{\theta \eta^{1/2}} \right) \right]}.
\label{eq:ES9}
\end{equation}
By multiplying $\frac{\kappa \eta}{\Pi^3 \phi}$ and we have the following form from Eq.~(\ref{eq:ES9}) 
\begin{equation}
\frac{2}{3} = \Pi^2 \theta \frac{\phi }{\kappa} \frac{1}{\eta^{1/2}}\left[1 - \exp \left( -\frac{\Pi}{\theta \eta^{1/2}} \right) \right].
\label{eq:ES10}
\end{equation}
In order to see the actual behavior of $\Pi$, I applied the third term perturbation method. As it belongs to the problem of the singular perturbation\cite{Holmes}, therefore here we assume
\begin{equation}
\Pi = \frac{1}{\varepsilon^{\gamma} } \left( \Pi_0 + \varepsilon^{\alpha} \Pi_1 + \varepsilon^{2\alpha} \Pi_2 + \cdots \right)
\label{eq:ES11}
\end{equation}
where $\gamma$ and $\alpha$ are constant, $\varepsilon = 1/ \theta \eta^{1/2}$.

By applying the Taylor expansion on the exponential part and substituting Eq.~(\ref{eq:ES11}) into Eq.~(\ref{eq:ES10}), we have
\begin{eqnarray}
\theta^2 \frac{\phi }{\kappa} \varepsilon \Pi^2 \left\{ \varepsilon \Pi  - \frac{1}{2}\varepsilon^2 \Pi^2 + \frac{1}{6}\varepsilon^3 \Pi^3 \cdots \right\} =  \frac{2}{3}  \nonumber   \\ 
\Leftrightarrow \theta^2 \frac{\phi }{\kappa} \varepsilon^{1-2 \gamma} \left( \Pi_0^2  + 2 \varepsilon^{\alpha} \Pi_1 \Pi_0 +  2 \varepsilon^{2 \alpha} \Pi_2 \Pi_0 +  \varepsilon^{2 \alpha} \Pi_1^2 + \cdots  \right) \{ \varepsilon^{1-\gamma} \left(\Pi_0 + \varepsilon^{\alpha} \Pi_1 + \varepsilon^{2\alpha} \Pi_2 + \cdots \right) \nonumber  \\  
- \frac{1}{2}\varepsilon^{2-2 \gamma} \left(\Pi_0^2  + 2 \varepsilon^{\alpha} \Pi_1 \Pi_0 +  2 \varepsilon^{2 \alpha} \Pi_2 \Pi_0 +  \varepsilon^{2 \alpha} \Pi_1^2 + \cdots \right) + \frac{1}{6}\varepsilon^{3-3 \gamma} \left(\Pi_0^3 \cdots \right)\cdots \} = \frac{2}{3} 
\label{eq:ES12}
\end{eqnarray}
as $\varepsilon \rightarrow 0$.

To balance each terms, we find that $\gamma=2/3$ and $\alpha=1/3$ then we obtain, 
\begin{eqnarray}
\theta^2 \frac{\phi }{\kappa}\left( \Pi_0^2  + 2 \varepsilon^{1/3} \Pi_1 \Pi_0 +  2 \varepsilon^{2/3} \Pi_2 \Pi_0 + \varepsilon^{2/3} \Pi_1^2 + \cdots  \right) \{\Pi_0 + \varepsilon^{1/3} \Pi_1 + \varepsilon^{2/3} \Pi_2 + \cdots \nonumber  \\  
- \frac{1}{2}\varepsilon^{1/3} \left(\Pi_0^2  + 2 \varepsilon^{1/3} \Pi_1 \Pi_0 +  2 \varepsilon^{2/3} \Pi_2 \Pi_0 + \varepsilon^{2/3} \Pi_1^2 + \cdots \right) + \frac{1}{6}\varepsilon^{2/3} \left(\Pi_0^3 \cdots \right)\cdots \} = \frac{2}{3}.   \label{eq:ES13}
\end{eqnarray}
From this we have
\begin{eqnarray}
O\left(1 \right) \Leftrightarrow \theta^2 \frac{\phi }{\kappa} \Pi_0^3 &=& \frac{2}{3} \nonumber \\ 
\Pi_0 &=& \left( \frac{2}{3} \right)^{\frac{1}{3}} \frac{1 }{\theta^{2/3}} \left( \frac{\kappa }{\phi}\right)^{\frac{1}{3}} 
\label{eq:ES14}
\end{eqnarray}
\begin{eqnarray}
O\left(\varepsilon^{1/3} \right) \Leftrightarrow 3\Pi_0^2 \Pi_1 - \frac{1}{2}\Pi_0^4 &=& 0 \nonumber \\
\Pi_1 &=& \frac{1}{6} \Pi_0^2 = \frac{1}{ \theta^{4/3}} \left( \frac{\kappa }{\phi} \right)^{\frac{2}{3}} \left(\frac{1}{486} \right)^{\frac{1}{3}}
\label{eq:ES15}
\end{eqnarray}
\begin{eqnarray}
O\left(\varepsilon^{2/3} \right) \Leftrightarrow 3\Pi_0^2 \Pi_2 - 2 \Pi_0^3\Pi_1 +\frac{1}{6} \Pi_0^5 +3 \Pi_0\Pi_1^2 &=& 0 \nonumber \\
\Pi_2 &=& \frac{2}{3} \Pi_0 \Pi_1   -\frac{1}{18} \Pi_0^3   - \frac{\Pi_1^2}{\Pi_0}= \frac{1}{54 \theta^2}  \frac{\kappa }{\phi}
\label{eq:ES16}
\end{eqnarray}
Using results of Eq.~(\ref{eq:ES14}), Eq.~(\ref{eq:ES15}), Eq.~(\ref{eq:ES16}), $\gamma = 2/3$ and $\alpha = 1/3$ for Eq.~(\ref{eq:ES11}) then we have a following result,
\begin{equation}
\Pi = \frac{ \kappa}{54 \phi \theta^2 } + \left(\frac{\kappa^2 }{486 \phi^2 \theta^{3} }\right)^{\frac{1}{3}} \eta^{\frac{1}{6}} + \left( \frac{2 \kappa}{3 \phi} \right)^{\frac{1}{3}}\eta^{\frac{1}{3}}
\label{eq:ES17}
\end{equation}
which corresponds to the Eq.~(7).

\section{Self-similarity of the first class and the second class for the data plots using the surface coated with grease}

In this paper, to focus on the viscoelastic behavior of the bulk and to eliminate the adhesion effect on the surface of PDMS board, chalk powder was dusted on the surface. Coating the surface with grease is expected to be the same effect though the plots are comparatively scattered particularly for the larger spheres. This may be due to the surface tension and that the surface interaction is not completely eliminated.
\begin{figure*}[t]
\begin{center}
\includegraphics[width=15cm]{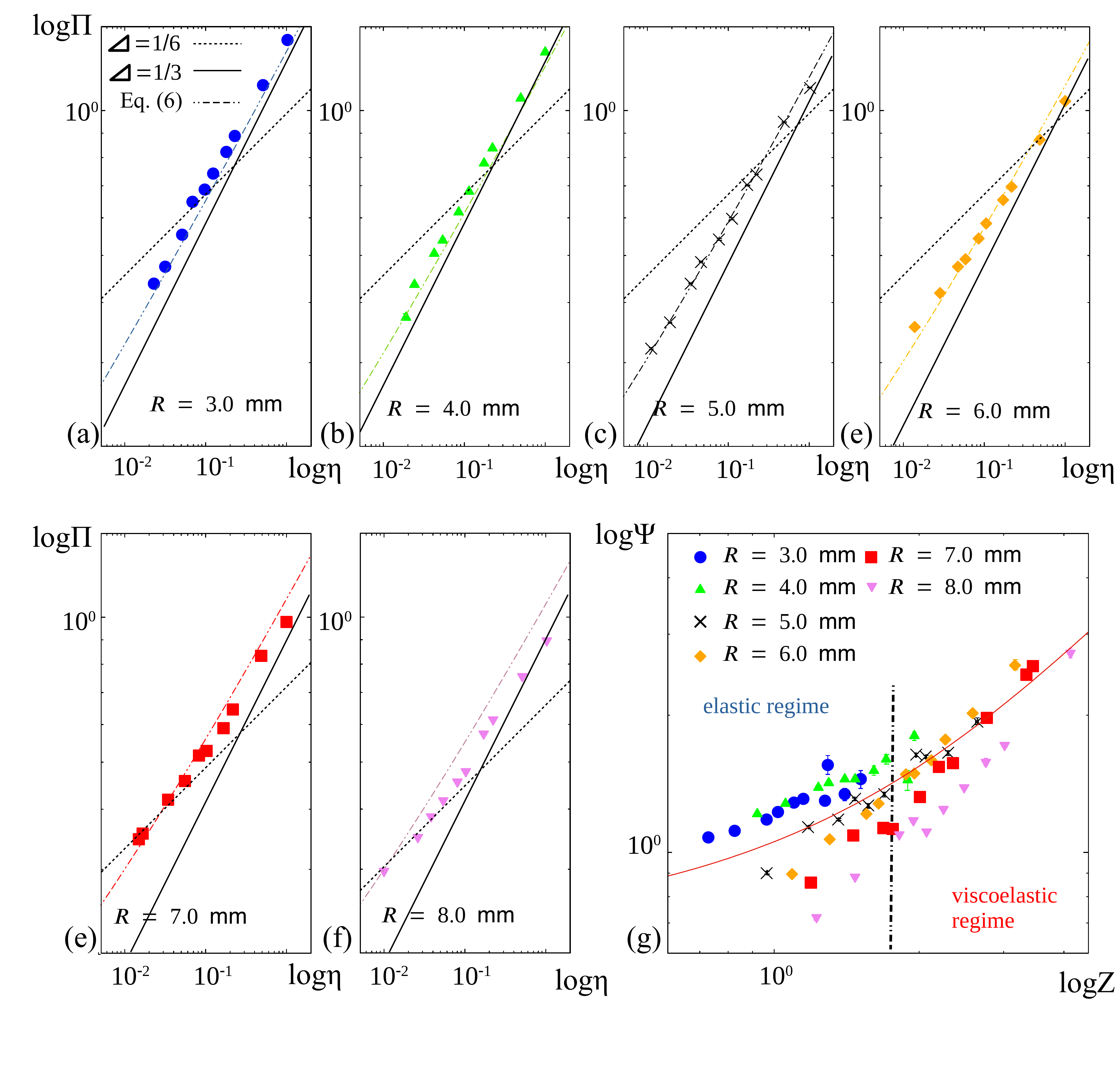}
\caption{(Color online) The different hierarchical structures of self-similarity for the experiments in which the PDMS surface is coated with grease (WD-40). (a) - (f) Self-similarity of the first class: the power law relations $\Pi$ and $\eta$ for spheres of different sizes. The dashed lines indicate the slope of 1/6, the solid lines indicate the slope of 1/3 and the colored dot-dashed line indicates Eq.~(7) for each size of the spheres. (g) Self-similarity of the second class: the plots between $\Psi$ and $Z$. $R = {\rm 3.0~mm}$ ($\textcolor{blue}{\bullet}$), ${\rm 4.0~mm}$ ($\textcolor{green}{\blacktriangle}$), {\rm 5.0~mm} ($\times$), {\rm 6.0~mm} ($\textcolor{orange}{\blacklozenge}$), {\rm 7.0~mm} ($\textcolor{red}{\blacksquare}$)  and {\rm 8.0~mm} ($\textcolor[rgb]{0.7, 0.4, 0.9}{\blacktriangledown}$) where $\Pi=\delta_m/R$, $\eta=\rho v_i^2/E$, $\Psi=\frac{\Pi^3 \phi}{\kappa \eta} = \frac{ \delta_m^3 E \phi}{ R^{2} h \rho v_i^2} $ and $Z = \frac{ \Pi}{ \theta \eta^{1/2}} = \frac{ E \delta_m }{ \mu  v_i}$. The red line in Figure (g) is Eq.~(6). The dashed line roughly indicates the line separating the region.}
\label{fig:FSS2}
\end{center}
\end{figure*} 

\section{The perturbation of the Kelvin-Voigt Viscoelastic Foundation model}

Here we show the behavior of dynamical impact for the case in which the Kelvin-Voigt model is applied for the each unit of foundations instead of the Maxwell model. In the Kelvin-Voigt model, the relation between the stress and deformation is described as $\sigma = E \epsilon + \mu \frac{d \epsilon}{dt} $. As it was assumed in the manuscript $\frac{d \delta}{dt} = v_i$, the exchange of energy for the maximum deformation is described as follows,
\begin{equation}
\frac{2}{3} \pi R^3 \rho v_i^2   = \frac{ \pi E \phi R \delta_m^3}{3 h } +  \frac{ \pi \mu \phi R \delta_m^2}{ h }\frac{d \delta}{dt}.
\label{eq:ES18}
\end{equation}
Introducing the dimensionless numbers in Eq. (4), we have the following dimensionless form from Eq. (\ref{eq:ES18}) 
\begin{equation}
\frac{2}{3}  = \frac{1}{3} \frac{\Pi^3 \phi}{\kappa \eta} +  \frac{\Pi^2 \theta \phi}{\kappa \eta^{1/2}}. 
\label{eq:ES19}
\end{equation}

In the same way as MVF model, I apply the perturbation of Eq.~(\ref{eq:ES11}) as follows,
\begin{equation}
\frac{1}{3}\theta^2 \frac{\phi }{\kappa} \varepsilon^{2-3\gamma}  \left( \Pi_0^3  + 3 \varepsilon^{2\alpha} \Pi_0^2 \Pi_1 + \cdots \right)    + \theta^2 \frac{\phi }{\kappa} \varepsilon^{1-2\gamma}  \left( \Pi_0^2  + 2 \varepsilon^{\alpha} \Pi_0 \Pi_1 + \cdots \right) = \frac{2}{3}
\label{eq:ES20}
\end{equation}
as $\varepsilon \rightarrow 0$.

Considering the balance, there is two possibility: $\gamma = \frac{2}{3}$ or  $\gamma=\frac{1}{2}$. However, $\gamma = \frac{2}{3}$ is impossible as $O\left(\varepsilon^{- \frac{1}{3}}\right)$ appears and it is not higher order of $O\left(1\right)$ while $\gamma = \frac{1}{2}$ is possible and it is well ordered. Therefore, we have $\gamma = \frac{1}{2}$, $\alpha = \frac{1}{2}$, then the solution of the perturbation is  as follows,
\begin{equation}
\Pi = \sqrt{\frac{2\kappa}{3\phi \theta} } \eta^{\frac{1}{4}}  - \frac{\kappa}{9 \phi\theta^2}.
\label{eq:ES21}
\end{equation}

It reveals 1/4 power-law on $\eta$. This solution is not consistent with our experimental observation and demonstrates that the crossover of scaling law does not occur. However, this solution is inconsistent with many points. Therefore, if the material obeys the Kelvin-Voigt model, it is possible that one cannot expect $\frac{d \delta}{dt} = v_i$. 

The self-similar solution by the variables of the second class will be as follows,
\begin{equation}
\Psi =\frac{2}{3}\frac{Z}{1+Z}
\label{eq:ES22}
\end{equation}
where $\Psi = \frac{\phi \Pi^3}{\kappa \eta}$ and $Z = \frac{\Pi}{\theta \eta^{1/2}}$. The elastic regime is recovered when $Z \rightarrow \infty$ though it cannot be achieved by decreasing velocity, $\eta \rightarrow 0$ as the perturbation result shows. It can be achieved when $\theta \rightarrow 0$. Therefore, the Kelvin-Voigt model cannot reveal crossover of scaling law.

\section{The perturbation of the Zener Viscoelastic Foundation model}

In this section, I show the result of the perturbation for the Zener Viscoelastic foundation model. The Zener model is a hybrid of the Maxwell model and the Kelvin-Voigt model. The unit of each foundation is described by 
\begin{equation}
\frac{\mu}{E}\frac{d \sigma}{dt} = -\sigma + E_K \epsilon + \mu \left(  \frac{E_K}{E}+1\right) \frac{d \epsilon}{dt}
\label{eq:ES23}
\end{equation}
where $E_K$ is the spring unit parallelly connected with its Maxwell element composed of a serial connection of a dashpot $\mu$ and a spring $E$.  Assuming $\frac{d \delta}{dt} = v_i$, the energy exchange will be as follows,
\begin{equation}
\frac{2}{3} \pi R^3 \rho v_i^2   = \frac{ \pi E_K \phi R \delta_m^3}{3 h } +  \frac{ \pi \mu \phi R \delta_m^2}{ h }\frac{d \delta}{dt}\left[ 1-\exp \left(- \frac{E t_c}{\mu}\right) \right] .
\label{eq:ES24}
\end{equation}
By applying dimensional analysis based on Eq. (4) and using $t_c = \frac{\delta_m}{v_i}$, dimensionless form will be 
\begin{equation}
\frac{2}{3}  = \frac{ \phi \Pi^3 }{3 \nu \kappa \eta } + \Pi^2 \theta \frac{\phi }{\kappa} \frac{1}{\eta^{1/2}}\left[1 - \exp \left( -\frac{\Pi}{\theta \eta^{1/2}} \right) \right]
\label{eq:ES25}
\end{equation}
where $\nu = \frac{E}{E_K}$.  Applying the perturbation of Eq.~(\ref{eq:ES11}) and Taylor expansion to the exponential part, the result will be as follows,
\begin{eqnarray}
\theta^2 \frac{\phi }{\kappa} \left\{  \frac{1}{3 \nu} \varepsilon^2 \Pi^3 +   \varepsilon \Pi^2\left( \varepsilon \Pi  - \frac{1}{2}\varepsilon^2 \Pi^2 + \frac{1}{6}\varepsilon^3 \Pi^3 \cdots  \right) \right\} =  \frac{2}{3}  \nonumber   \\ 
\Leftrightarrow \theta^2 \frac{\phi }{\kappa} [ \frac{1}{3\nu} \varepsilon^{2-3\gamma} \left\{ \Pi_0^3 +3\varepsilon^{\alpha}\Pi_0^2 \Pi_1 + \left( 3 \Pi_0^2 \Pi_2 + 3 \Pi_0 \Pi_1^2 \right) \varepsilon^{2 \alpha}   + \cdots  \right\} +   \nonumber  \\  
\varepsilon^{1-2 \gamma} \left( \Pi_0^2  + 2 \varepsilon^{\alpha} \Pi_1 \Pi_0 +  2 \varepsilon^{2 \alpha} \Pi_2 \Pi_0 +  \varepsilon^{2 \alpha} \Pi_1^2 + \cdots  \right) \{ \varepsilon^{1-\gamma} \left(\Pi_0 + \varepsilon^{\alpha} \Pi_1 + \varepsilon^{2\alpha} \Pi_2 + \cdots \right)   \nonumber  \\  
- \frac{1}{2}\varepsilon^{2-2 \gamma} \left(\Pi_0^2  + 2 \varepsilon^{\alpha} \Pi_1 \Pi_0 +  2 \varepsilon^{2 \alpha} \Pi_2 \Pi_0 +  \varepsilon^{2 \alpha} \Pi_1^2 + \cdots \right) + \frac{1}{6}\varepsilon^{3-3 \gamma} \left(\Pi_0^3 \cdots \right)\cdots \} ] = \frac{2}{3} 
\label{eq:ES26}
\end{eqnarray}
as $\varepsilon \rightarrow 0$.

To balance each terms, $\gamma = \frac{2}{3}$, $\alpha = \frac{1}{3}$. By considering the orders $O\left(1\right)$, $O\left(\varepsilon^{\frac{1}{3}} \right)$ and $O\left(\varepsilon^{\frac{2}{3}} \right)$ to determine $\Pi_0$, $\Pi_1$ and $\Pi_2$, we have the following perturbation result,
\begin{equation}
\Pi = \left(3K^3-2K^2\right) \frac{ \kappa}{54 \phi \theta^2 } + K^{\frac{5}{3}}\left(\frac{\kappa^2 }{486 \phi^2 \theta^{3} }\right)^{\frac{1}{3}} \eta^{\frac{1}{6}} +K^{\frac{1}{3}} \left( \frac{2 \kappa}{3 \phi} \right)^{\frac{1}{3}}\eta^{\frac{1}{3}}
\label{eq:ES27}
\end{equation}
where $K = \frac{3 \nu}{3\nu +1}$. 

As $E \gg E_K $ recovers the Maxwell model, it corresponds to $\nu \gg 1$ and $K \rightarrow 1$. In this case, we have Eq.~(\ref{eq:ES17}), which is the same result of the MVF model.

The self-similar solution is, using $\Psi = \frac{\phi \Pi^3}{\kappa \eta}$ and $Z = \frac{\Pi}{\theta \eta^{1/2}}$
\begin{equation}
\Psi =\frac{2 \nu Z}{Z + 3 \nu \left[1- \exp \left( -Z \right)\right]}.
\label{eq:ES28}
\end{equation}
Eq. (\ref{eq:ES28}) corresponds to Eq. (6) by $\nu \rightarrow \infty$, which is the limit of Maxwell. As Eq. (\ref{eq:ES27}) shows, the Zener Viscoelastic Foundation model reveals the crossover of scaling law depending on the impact-velocity. The only difference from MVF model is just the coefficient of  $K = \frac{3 \nu}{3\nu +1}$. However, there is no perturbation result for the limit of the Kelvin-Voigt, which is realized by  $E \ll E_K $, as it gives $K \rightarrow 0$ then the solution vanishes. This consideration suggests that the Maxwell element, the serial connection of spring and dashpot, is essential to realize the crossover of scaling law.
